\documentclass[prd,aps,onecolumn,superscriptaddress,nofootinbib,showpacs,preprintnumbers,longbibliography, dvipsnames]{revtex4-1}

\UseRawInputEncoding
\usepackage[T1]{fontenc} 
\usepackage[utf8]{inputenc}

\usepackage{float} 
\usepackage{feynmp-auto} 
\usepackage{tensor} 
\usepackage{physics} 
\usepackage{subfig}
\usepackage{listings}
\usepackage{url}
\usepackage{slashed}
\usepackage{ulem}
\usepackage{hyperref}
\hypersetup{colorlinks=true, linkcolor=teal, urlcolor=blue, citecolor=red}

\usepackage{adjustbox}

\usepackage{tikz}
\usetikzlibrary{shapes.geometric,arrows}

\usepackage{amsmath}
\usepackage{amssymb}
\usepackage{bm}
\usepackage{bbold}

\usepackage{graphicx}
\usepackage[american]{babel}
\usepackage{enumerate}
\usepackage{mathtools}

\usepackage{multirow}  
\usepackage{makecell} 
\usepackage{booktabs}

\usepackage{cancel}

\usepackage{xcolor}

\newcommand{\U}[2]{\mathrm{U}(#1)_{\mathrm{#2}}}		
\newcommand{\SU}[2]{\mathrm{SU}(#1)_{\mathrm{#2}}}		
\newcommand{\E}[1]{\mathrm{E}_{#1}}	

\renewcommand{\[}{\left[}

\usepackage{pifont}

\newcommand\varpm{\mathbin{\vcenter{\hbox{%
  \oalign{\hfil$\scriptstyle\hspace{-0.2ex}+\hspace{-0.2ex}$\hfil\cr
          \noalign{\kern-.5ex}
          $\scriptscriptstyle({-})$\cr}%
}}}}

\newcommand\varmp{\mathbin{\vcenter{\hbox{%
  \oalign{\hfil$\scriptstyle\hspace{-0.2ex}-\hspace{-0.2ex}$\hfil\cr
          \noalign{\kern-.5ex}
          $\scriptscriptstyle({+})$\cr}%
}}}}

\definecolor{bostonuniversityred}{rgb}{0.8, 0.0, 0.0}

\newcommand{\green}[0]{\color{ForestGreen}}

\usepackage[capitalise]{cleveref}

\crefname{section}{Sec.}{Secs.}
\crefname{table}{Tab.}{Tabs.}
\crefname{figure}{Fig.}{Figs.}
\crefname{equation}{Eq.}{Eqs.}
\crefname{appendix}{Appendix\ }{Appendix\ }

\allowdisplaybreaks

\begin{document}

\begin{flushright}
CERN-TH-2022-217\\
\vskip1cm
\end{flushright}
	
	\title{Deep Learning Searches for Vector-Like Leptons at the LHC \\ and Electron/Muon Colliders}
	\author{Ant\'onio P. Morais} 
		\email{aapmorais@ua.pt}
		\affiliation{
		\sl Departamento de F\'{i}sica da Universidade de Aveiro and Centre for Research and Development in Mathematics and Applications (CIDMA), 3810-183 Aveiro, Portugal \\
		}
	    \affiliation{
		\sl Theoretical Physics Department, CERN, 1211 Geneva 23, Switzerland \\
		}
	\author{Ant\'onio Onofre} 
	    \email{onofre@fisica.uminho.pt}
		\affiliation{
		\sl Departamento de F\'{i}sica da Universidade do Minho, 4710-057 Braga, Portugal \\
		}
	\author{Felipe F. Freitas} 
	    \email{felipefreitas@ua.pt}
	    \affiliation{
		\sl Departamento de F\'{i}sica da Universidade de Aveiro and Centre for Research and Development in Mathematics and Applications (CIDMA), 3810-183 Aveiro, Portugal \\
		}
	\author{Jo\~ao Gon\c{c}alves} 
	    \email{jpedropino@ua.pt}
	\affiliation{
		\sl Departamento de F\'{i}sica da Universidade de Aveiro and Centre for Research and Development in Mathematics and Applications (CIDMA), 3810-183 Aveiro, Portugal \\
		}
	\author{Roman Pasechnik} 
	    \email{Roman.Pasechnik@thep.lu.se}
	    \affiliation{
		\sl Department of Astronomy and Theoretical Physics, Lund University, S\"{o}lvegatan 14A, SE 223-62 Lund, Sweden \\
		}
	\author{Rui Santos} 
        \email{rasantos@fc.ul.pt}
	    \affiliation{
		\sl Centro de F\'{i}sica Te\'orica e Computacional, Faculdade de Ci\^{e}ncias, Universidade de Lisboa, 1749-016 Lisboa, Portugal \\
		}
		\affiliation{
		\sl ISEL - Instituto Superior de Engenharia de Lisboa, Instituto Polit\'ecnico de Lisboa, 1959-007 Lisboa, Portugal \\
		}
	
	
	\begin{abstract}
		The discovery potential of both singlet and doublet vector-like leptons (VLLs) at the Large Hadron Collider (LHC) as well as at the not-so-far future muon and electron machines is explored. The focus is on a single production channel for LHC direct searches while double production signatures are proposed for the leptonic colliders. A Deep Learning algorithm to determine the discovery (or exclusion) statistical significance at the LHC is employed. While doublet VLLs can be probed up to masses of 1 TeV, their singlet counterparts have very low cross sections and can hardly be tested beyond a few hundreds of GeV at the LHC. This motivates a physics-case analysis in the context of leptonic colliders where one obtains larger cross sections in VLL double production channels, allowing to probe higher mass regimes otherwise inaccessible even to the LHC high-luminosity upgrade.
 	\end{abstract}
	
	\maketitle

\section{Introduction}\label{sec:Introduction}

The Standard Model (SM) of particle physics has, so far, served as the guide towards unraveling the nature of all subatomic phenomena and its success is now indisputable. However, it is clear that it lacks some of the necessary ingredients to fully describe nature, such as an explanation for neutrino masses, as firstly observed in neutrino oscillations by experiments like the Super-Kamiokande \cite{Fukuda:1998mi} or a particle candidate to explain experimental evidences for the existence of Dark Matter (DM) \cite{Bertone:2004pz}. 
Recently, other discrepancies have also come into the forefront, such as the measurement of the anomalous magnetic moment of the muon at Fermilab \cite{Abi:2021gix}, indicating a tension of $4.2\sigma$ with the SM predictions, as well as the recent results coming from the LHCb experiment, where hints of lepton flavour universality violation have been reported at a $3.1\sigma$ significance \cite{Aaij:2021vac}. Naturally, such deviations do not meet the $5\sigma$ requirement to indicate the existence of new physics (NP) and more data still needs to be collected if the experimental deviations are confirmed. Together with the theoretical calculations in the framework of the SM, an explanation of anomalies requires the addition of NP.

Finding well motivated NP scenarios that simultaneously address all the aforementioned questions is of utmost importance and, in particular, Grand Unified theories (GUTs) provide interesting avenues to follow. These classes of models typically predict new particles present at the TeV scale (or not too far from it), that may be probed at current and future colliders. Therefore, the study of simplified frameworks that can later be mapped to the low-energy limit of a certain UV complete framework can help in constraining their viable parameter space. In particular, some of the authors have proposed a GUT framework, which unifies all matter, Higgs and gauge sectors within the $\E{8}$ group, either at a conventional high-scale scenario \cite{Morais:2020ypd,deAnda:2020prd}, or a low-scale one \cite{Aranda:2021bvg,Aranda:2021eyn}. It turns out that, as a common prediction of any of such proposals, the presence of new doublet Vector-Like Lepton (VLL) particles at the TeV scale or below is expected. A phenomenological analysis of doublet type VLLs, in the context of GUT model, was extensively discussed by some of the authors in a previous work \cite{Freitas:2020ttd}. Naturally, such description was bounded by the constraints and details of the underlying UV theory and as such, the main VLL features can only be fully captured in more simple phenomenological models, without the vast particle content and constraints of GUT frameworks. Additionally, singlet-type VLLs, which are not predicted in the context of the previously considered GUT, can also be constructed and represent a significant par of this work. Studying the differences between both types of VLL is of relevance not only in the context of the LHC but also at future proposed leptonic colliders. Furthermore, if the anomaly in the measurement of the magnetic moment of the muon is confirmed, which can happen in about one year from the date this article is being written, a possible explanation can arise in the form of TeV scale VLLs. In this context, we refer to \cite{Yin:2020afe,Crivellin:2020ebi,Bissmann:2020lge,Hiller:2019mou,Hiller:2020fbu,Guedes:2021oqx} for further recent VLLs phenomenological studies.

With the above arguments in mind we study the phenomenological viability of simple extensions of the SM enlarged with a single VLL and a right-handed neutrino. For completeness, we consider two scenarios, the $\mathrm{SU(2)_L}$ doublet and the singlet VLLs. Due to their vector-like nature, they escape constraints from the fourth generation searches \cite{Eberhardt:2012gv}, while non-vanishing couplings to the SM can induce contributions not only to the muon $(g-2)_\mu$ anomaly, but also opening tightly constrained channels such as $\mu \rightarrow e\gamma$ \cite{Zyla:2020zbs}.

This article is organized as follows. In \cref{sec:SM_plus_VLL}, we introduce the simplified models, presenting the relevant theoretical details that will serve as the basis behind the subsequent numerical simulations. In \cref{sec:collider_and_g2} we introduce the numerical methodology for the studies we conduct in this work. In particular, in \cref{subsec:Muon_g2}, we discuss the impact of the main flavour constraints associated with the addition of a VLL, while in \cref{subsec:collider_methods} we delve into collider analysis, both in the context of the LHC and future electron (linear) or muon colliders. Finally, in \cref{sec:Results} we present our main results and conclude in \cref{sec:Conclusions}.

\section{The models}
\label{sec:SM_plus_VLL}

In this article, we study collider phenomenology of both $\mathrm{SU(2)_L}$ doublet and singlet VLL extensions of the SM accompanied with new right-handed neutrinos. We present the NP and SM fields' quantum numbers in \cref{tab:VLL_RN,tab:SM-numbers} respectively.
\begin{table}[H]
	\centering
	\begin{tabular}{c|c|c|c}
		\textbf{Field} & $\mathbf{SU(3)_\text{C}}$ & $\mathbf{SU(2)_\text{L}}$ & $\mathbf{U(1)_\text{Y}}$ \\ \hline
		$E_\mathrm{L,R}$          & \textbf{1}                & \textbf{2}                & $-1/2$                            \\
		$\mathcal{E}_\mathrm{L,R}$          & \textbf{1}                & \textbf{1}                & $1$                         \\
		$\nu_\mathrm{R}$          & \textbf{1}                & \textbf{1}                & $0$                           
	\end{tabular}
	\caption{\label{tab:VLL_RN} Quantum numbers of the new exotic fermions. We adopt the notation of $E_{\mathrm{L,R}}$ being an $\mathrm{SU(2)}$ doublet, while $\mathcal{E}_{\mathrm{L,R}}$ is a singlet, and $\nu_\mathrm{R}$ is the new right-handed neutrino.}
\end{table}

\begin{table}[H]
	\centering
	\begin{tabular}{c|c|c|c|c}
		\textbf{Field} & $\mathbf{SU(3)_\text{C}}$ & $\mathbf{SU(2)_\text{L}}$ & $\mathbf{U(1)_\text{Y}}$ & \textbf{\# of generations} \\ \hline
		$Q_\mathrm{L}$          & \textbf{3}                & \textbf{2}                & $1/6$   & 3                          \\
		$L$            & \textbf{1}                & \textbf{2}                & $-1/2$  & 3                          \\
		$d_\mathrm{R}$          & \textbf{3}                & \textbf{1}                & $-1/3$  & 3                          \\
		$u_\mathrm{R}$          & \textbf{3}                & \textbf{1}                & $2/3$   & 3                          \\
		$e_\mathrm{R}$          & \textbf{1}                & \textbf{1}                & $-1$              & 3 \\
		$\phi$          & $\bm{1}$                & $\bm{2}$                & $1/2$   & 1
	\end{tabular}
	\caption{\label{tab:SM-numbers} Quantum numbers for the SM fields. Here, $Q_{\mathrm{L}}$, $d_\mathrm{R}$, $u_\mathrm{R}$ are the quark fields, $L$ and $e_\mathrm{R}$ are the lepton fields, and $\phi$ is the Higgs doublet.}
\end{table}
The $\mathrm{SU(2)_L}$ doublets are defined as
\begin{equation}\label{eq:doublets}
Q_L^i = \begin{bmatrix}
u_{\mathrm{L}} \\ d_{\mathrm{L}}
\end{bmatrix}^i, \quad L^i = \begin{bmatrix}
\nu_{\mathrm{L}} \\ e_{\mathrm{L}}
\end{bmatrix}^i, \quad E_{\mathrm{L,R}} = \begin{bmatrix}
\nu'_{\mathrm{L,R}} \\ e'_{\mathrm{L,R}}
\end{bmatrix},\quad
\phi = \begin{bmatrix}
\phi^+ \\ \phi^0
\end{bmatrix},
\end{equation}
where $i$ is a generation index, $i = 1,2,3$. The most general and renormalizable Lagrangian density of the doublet VLL model, consistent with the symmetries in Tab.~\ref{tab:VLL_RN}, reads as
\begin{equation}\label{eq:Doubet_case}
\begin{aligned}
\mathcal{L}_d = &\qty(\Theta_i  \bar{E}_{\mathrm{L}} e_{\mathrm{R}}^i \phi + \Upsilon \bar{E}_{\mathrm{L}} \nu_{\mathrm{R}} \tilde{\phi} + \Sigma_i  \bar{L}^i \nu_{\mathrm{R}} \tilde{\phi} + \Omega \bar{E}_{\mathrm{R}}\nu_R \tilde{\phi} + \Pi_{ij} \bar{L}^i e_{\mathrm{R}}^j\phi + \mathrm{H.c.}) + \\ & + M_{\rm E} \bar{E}_{\mathrm{L}} E_{\mathrm{R}} + M_{\rm LE}^i \bar{L}_i E_{\mathrm{R}} + \frac{1}{2} M_{\nu_{\mathrm{R}}}\bar{\nu}_R \nu_R \,,
\end{aligned}
\end{equation}
where the Yukawa couplings are denoted by $\Theta$, $\Upsilon$, $\Sigma$, $\Omega$ and $\Pi$, whereas bi-linear mass terms are indicated by $M_E$, $M_{LE}$ and $M_{\nu_{\mathrm{R}}}$.
For the singlet VLL case, the Lagrangian is written as 
\begin{equation}\label{eq:Singlet_case}
\begin{aligned}
\mathcal{L}_s = &\qty(\theta_i\bar{L}^i \mathcal{E}_{\mathrm{R}}\phi + \sigma^i \bar{L}_i \nu_{\mathrm{R}} \tilde{\phi} + \pi_{ij} \bar{L}^i e_{\mathrm{R}}^j\phi +\mathrm{H.c.}) +  M_{\rm E} \bar{\mathcal{E}}_{\mathrm{L}} \mathcal{E}_{\mathrm{R}}  + \frac{1}{2} M_{\nu_{\mathrm{R}}}\bar{\nu}_R \nu_R \,,
\end{aligned}
\end{equation}
where $\pi$, $\sigma$ and $\theta$ are the corresponding Yukawa couplings. The mass matrices for both the singlet and the doublet scenarios, expressed in the $\{e_L^1, e_L^2, e_L^2, e_L'\} \otimes \{e_R^1, e_R^2, e_R^2, e_R'\}$ basis, take the following form
\begin{equation}\label{eq:Mass_matrices}
M_{L}^{\mathrm{doublet}} = \begin{bmatrix}
\qty(\dfrac{v \Pi}{\sqrt{2}})_{3\times 3} & \qty(M_{\rm LE})_{3\times 1} \\[1.5em]
\qty(\dfrac{v \Theta}{\sqrt{2}})_{1\times 3} & \qty(M_{\rm E})_{1\times 1}
\end{bmatrix},
\quad 
M_{L}^{\mathrm{singlet}} = \begin{bmatrix}
\qty(\dfrac{v \pi}{\sqrt{2}})_{3\times 3} & \qty(\dfrac{v \theta}{\sqrt{2}})_{3\times 1} \\[1.5em]
\qty(0)_{1\times 3} & \qty(M_{\rm E})_{1\times 1}
\end{bmatrix} \,,
\end{equation}
with $v = 246~\mathrm{GeV}$ the electroweak symmetry breaking (EWSB) Higgs doublet VEV $\left( 0,~v \right)^\top$, and
whose eigenstates describe both the chiral SM-like charged leptons as well as the new VLLs. 

For the neutrino sector the mass matrices can be expressed as
\begin{equation}\label{eq:Matrices_neutrinos}
\begin{aligned}
M_{\nu}^{\mathrm{doublet}} =
\begin{bmatrix}
\qty(0)_{3\times 3} & \qty(\dfrac{\sigma v}{\sqrt{2}})_{3\times 1} & \qty(0)_{3\times 1} & \qty(M_{\rm LE})_{3\times 1} \\
\qty(\dfrac{\sigma v}{\sqrt{2}})_{1\times 3} & M_{\nu_\mathrm{R}} & \qty(\dfrac{\Upsilon v}{\sqrt{2}})_{1\times 1} & \qty(\dfrac{\Omega v}{\sqrt{2}})_{1\times 1} \\
\qty(0)_{1\times 3} & \qty(\dfrac{\Upsilon v}{\sqrt{2}})_{1\times 1} & (0)_{1\times 1} & (M_{\rm E})_{1\times 1}& \\
\qty(M_{\rm LE})_{1\times 3} & \qty(\dfrac{\Omega v}{\sqrt{2}})_{1\times 1} & (M_{\rm E})_{1\times 1} & \qty(0)_{1\times 1} 
\end{bmatrix}\,,
\qquad
M_{\nu}^{\mathrm{singlet}} = \begin{bmatrix}
(0)_{3\times 3} & \qty(\dfrac{\sigma v}{\sqrt{2}})_{3\times 1} \\
\qty(\dfrac{\sigma v}{\sqrt{2}})^\dagger_{1\times 3} & \qty(M_{\nu_{\mathrm{R}}})_{1\times 1}
\end{bmatrix} \,.
\end{aligned}
\end{equation}
where the $\{\nu_L^1, \nu_L^2, \nu_L^3, \nu_R, \nu_L^{'}, \nu_R^{'}\} \otimes \{\nu_L^1, \nu_L^2, \nu_L^3, \nu_R, \nu_L^{'}, \nu_R^{'}\}$ basis was used for the doublet scenario whereas that of the singlet model was chosen as $\{\nu_L^1, \nu_L^2, \nu_L^3, \nu_R\} \otimes \{\nu_L^1, \nu_L^2, \nu_L^3, \nu_R\}$.

In this article, one considers that the lightest beyond-the-SM (BSM) neutrino is in the KeV range and sterile, acting as missing energy in the detector. Therefore, while the neutrino mass matrices above have the standard type-I seesaw form, such a light sterile neutrino implies that $M_{\nu_{\mathrm{R}}}$ is equally small, which means that $\Sigma$, $\Upsilon$ and $\Omega$ (for the doublet scenario) as well as $\sigma$ (for the singlet scenario), need to be very small in order for both models to be consistent with the active neutrinos mass scale. While a detailed discussion for the neutrino mass generation mechanism is beyond the scope of this article, it is not a difficult task to minimally extend both the doublet and singlet scenarios, e.g.~by introducing the dimension-five Weinberg operator, without affecting our analysis and conclusions \cite{Strumia:2006db}. However, the fact that $\Sigma$, $\Upsilon$, $\Omega$ and $\sigma$ need to be very small is calling for the introduction of an approximate global lepton number symmetry, which, for convenience, we define as $\mathrm{U}(1)_e \times \mathrm{U}(1)_\mu \times \mathrm{U}(1)_\tau$, such that only leptons transform non-trivially according to the following quantum numbers:
\begin{table}[H]
	\centering
	\begin{tabular}{c|c|c|c}
		\textbf{Field} & $\mathrm{U(1)_e}$ & $\mathrm{U(1)_\mu}$ & $\mathrm{U(1)_\tau}$ \\ \hline
		$E_\mathrm{L,R}$          & $0$  & $1$    & $0$                             \\
		$\mathcal{E}_\mathrm{L,R}$          & $0$  & $1$    & $0$                          \\
		$\nu_\mathrm{R}$          & $0$  & $0$    & $0$     \\                     $L^1$          & $1$  & $0$    & $0$                          \\
		$L^2$          & $0$  & $1$    & $0$                         \\
		$L^3$          & $0$  & $0$    & $1$                         \\
		$e_\mathrm{R}^1$          & $1$  & $0$    & $0$                         \\ 
		$e_\mathrm{R}^2$          & $0$  & $1$    & $0$                          \\
		$e_\mathrm{R}^3$          & $0$  & $0$    & $1$                        
	\end{tabular}
	\caption{\label{tab:U3} Quantum numbers under the approximate lepton number symmetry.}
\end{table}
Such an approximate symmetry constrains the model parameters as follows:
\begin{itemize}
    \item $\Sigma$, $\Upsilon$, $\Omega$ (for the doublet case) and $\sigma$ (for the singlet case) must be tiny in consistency with the smallness of the active neutrino masses;
    \item the charged lepton-sector Yukawa couplings $\Pi$ (for the doublet model) and $\pi$ (for the singlet model), are approximately diagonal and off-diagonal elements can be ignored in our analysis;
    \item The VLLs solely couple to muons such that only $\Theta_2$, $M_{{\rm LE},2}$ (for the doublet scenario) and $\theta_2$ (singlet scenario) are sizeable and relevant for our numerical studies.
\end{itemize}

In what follows, we are essentially interested in lepton couplings to both gauge and Higgs bosons. In particular, the focus is on the $W\nu E(\mathcal{E})$, $Z^0E(\mathcal{E})\ell$ and $H E(\mathcal{E})\ell$ vertices. All numerical computations are performed in the mass basis such that one needs to rotate the gauge eigenstates via the bi-unitary transformations
\begin{equation}\label{eq:rotation_mass}
\begin{aligned}
&\mathrm{Leptons:}\hphantom{..} M_L^{\mathrm{diag}} = \qty(U_L^e)^\dagger M_L \qty(U_R^e), \\
&\mathrm{Neutrinos:}\hphantom{..} M_\nu^{\mathrm{diag}} = \qty(U_\nu)^\dagger M_\nu \qty(U_\nu).
\end{aligned}
\end{equation}
The list of all Feynman rules that are relevant to the numerical analysis are shown in Appendix~\ref{app:Feyn_rules}.

Before presenting the methodology behind the numerical calculations, it is instructive to revisit the main experimental constraints for VLLs as well as the most relevant couplings for this discussion. First, let us recall that there are not many direct searches for VLLs, at least, in comparison with their quark counterparts. In fact, LEP searches have lead to a lower bound on the VLLs mass of $M_{\mathrm{VLL}} > 100.6$ GeV \cite{Achard:2001qw}, while the most restrictive constraints are coming from CMS where direct searches for doublet VLLs with sizeable couplings to taus are excluded at 95\% Confidence Level (CL) in the 120-790 GeV mass range \cite{Sirunyan:2019ofn}. Additionally, there are searches for charged lepton resonances which constrain the mass of a hypothetical VLL to be above a few hundred GeV, depending on the underlying assumptions (114–176 GeV for VLL and 100–468 GeV for a heavy lepton in the type-III inverse seesaw model \cite{ATLAS:2015qoy}, as well as 80-210 GeV in Ref.~\cite{CMS:2012ra}). However, for the model under consideration, the $[\mathrm{U}(1)]^3$ lepton number symmetry dictates that the relevant couplings are those to muons such that the most restrictive search in \cite{Sirunyan:2019ofn} does not apply.

The left-mixing matrix that rotates from the flavour to the mass basis can be parametrized by a single mixing angle, $\alpha$, as follows
\begin{equation}\label{eq:mixng_alpha}
U_L^e = \begin{bmatrix}
1 & 0 & 0 & 0 \\
0 & \cos{\alpha} & 0 & -\sin{\alpha} \\
0 & 0 & 1 & 0 \\
0 & -\sin{\alpha} & 0 & -\cos{\alpha}
\end{bmatrix}.
\end{equation}
With the $\mathrm{U}(1)_e \times \mathrm{U}(1)_\mu \times \mathrm{U}(1)_\tau$ approximate symmetry the mass matrices for both the doublet and singlet cases take a simple form
\begin{equation}\label{eq:lepton_matrices}
M_L^{\mathrm{doublet}} = \begin{bmatrix}
\dfrac{v\Pi_{11}}{\sqrt{2}} & 0 & 0 & 0 \\
0 & \dfrac{v\Pi_{22}}{\sqrt{2}} & 0 & M_{LE, 2} \\
0 & 0 & \dfrac{v\Pi_{33}}{\sqrt{2}} & 0 \\
0 & \dfrac{v\Theta_{2}}{\sqrt{2}} & 0 & M_E
\end{bmatrix}\,,
\quad\quad 
M_L^{\mathrm{singlet}} = \begin{bmatrix}
\dfrac{v\pi_{11}}{\sqrt{2}} & 0 & 0 & 0 \\
0 & \dfrac{v\pi_{22}}{\sqrt{2}} & 0 & \dfrac{v\theta_{2}}{\sqrt{2}} \\
0 & 0 & \dfrac{v\pi_{33}}{\sqrt{2}} & 0 \\
0 & 0 & 0 & M_E
\end{bmatrix}\,.
\end{equation}
Besides lepton mixing terms one also considers a neutrino mixing with the following structure,
\begin{equation}\label{eq:neutrino_mixing}
\begin{aligned}
&U_\nu^{\mathrm{doublet}} = \begin{bmatrix}
(U_\mathrm{PMNS})_{11} & (U_\mathrm{PMNS})_{12} & (U_\mathrm{PMNS})_{13} & 0 & 0 & 0 \\
(U_\mathrm{PMNS})_{21} & (U_\mathrm{PMNS})_{22} & (U_\mathrm{PMNS})_{23} & 0 & 0 & 0 \\
(U_\mathrm{PMNS})_{31} & (U_\mathrm{PMNS})_{32} & (U_\mathrm{PMNS})_{33} & 0 & 0 & 0 \\
0 & 0 & 0 & 1 & 0 & 0 \\
0 & 0 & 0 & 0 & (U_\nu)_{55} & (U_\nu)_{56} \\
0 & 0 & 0 & 0 & (U_\nu)_{65} & (U_\nu)_{66} \\
\end{bmatrix}, \\[0.5em]
&U_\nu^{\mathrm{singlet}} = \begin{bmatrix}
(U_\mathrm{PMNS})_{11} & (U_\mathrm{PMNS})_{12} & (U_\mathrm{PMNS})_{13}   & 0\\
(U_\mathrm{PMNS})_{21} & (U_\mathrm{PMNS})_{22} & (U_\mathrm{PMNS})_{23} & 0 \\
(U_\mathrm{PMNS})_{31} & (U_\mathrm{PMNS})_{32} & (U_\mathrm{PMNS})_{33} & 0 \\
0 & 0 & 0 & 1
\end{bmatrix}
\end{aligned}
\end{equation}
where the $3\times3$ SM block is fixed by the Pontecorvo–Maki–Nakagawa–Sakata (PMNS) matrix \cite{Esteban:2020cvm}. Technically, and considering only such a block, the PMNS matrix must in rigour be defined as $U_\mathrm{PMNS} = U_\nu \cdot U_L^\dagger$ such that the actual mixing is instead given as $(U_\mathrm{PMNS})_{i1} = (U_\nu)_{i1}$, $(U_\mathrm{PMNS})_{i3} = (U_\nu)_{i3} $ and $(U_\mathrm{PMNS})_{i2} = (U_\nu)_{i2} \cos \alpha$ with $i = 1,2,3$. However, the benchmark scenarios considered in \cref{sec:Results} are such that $\cos \alpha \approx 1$ for both the doublet and the singlet models and therefore the parametrization in \cref{eq:neutrino_mixing} is consistent with a realistic lepton mixing.

The $2\times 2$ block in the doublet case represents the mixing between the $\nu_5$ and $\nu_6$ mass eigenstates. These elements do not contribute to the interaction vertices used in our analysis and therefore their size and numerical values are not relevant for the discussion. For consistency, we assume a generic $\mathcal{O}(1)$ mixing. It also follows from the approximate lepton number symmetry that the mixing between the right-handed neutrino and the remaining neutral leptons is negligible and can be ignored. This means that $(U_\nu)_{44} = 1$, $(U_\nu)_{4j} = (U_\nu)_{j4} = 0$ and the couplings between the SM and BSM neutrinos can be neglected. 

Let us note that if one requires consistency with the recently observed muon $(g-2)_\mu$ anomaly \cite{Abi:2021gix}, the preferred values of the cosine of the mixing angle $\alpha$ must not be far from unity. On the other hand, a small $\alpha$ (or equivalently, small Yukawa parameters $\theta_2$ and $\Theta_2$) is preferable in order to keep tree-level modifications to the Feynman rules for the $H\mu^+\mu-$ and $\mathrm{Z^0}\mu^+\mu^-$ vertices under control (a list of the rules are shown in appendix \ref{app:Feyn_rules}). Indeed, a large mixing angle would make the values of the branching fractions $\mathrm{BR}(Z^0\rightarrow \mu^+\mu^-)$ and $\mathrm{BR}(H\rightarrow \mu^+\mu^-)$ unacceptably large. Furthermore, and noting that the VLL only couples to muons, one can see from \cite{delAguila:2008pw} that EW global fits constrain the mixing angle to be less than 0.034, which we impose in our numerical analysis. With such a small mixing, lepton flavour violation observables can be kept under control and the VLL single production cross-section is, for the singlet model, dominated by the $W\nu_\ell \mathcal{E}$ vertex and, for the doublet model, by the $W\nu_L^\prime E$ vertex, as we discuss in \cref{subsec:collider_methods}. It follows from the mixing structure in \cref{eq:neutrino_mixing} (see also the Feynman rules in \cref{eq:Feynman_Rules_Wplus,eq:Feynman_Rules_Wminus,eq:Feynman_Rules_Wplus_1,eq:Feynman_Rules_Wminus_2}) that interactions involving right-handed neutrinos are absent in both scenarios. This is consistent with the approximate lepton number symmetry, where one can assume that the smallness of the $\sigma$ and $\Sigma$ Yukawa couplings, results in a negligible $W\nu_R \mathcal{E}$ and  $W\nu_R E$ interaction strength. Therefore, in our numerical analysis, one can ignore such interactions by safely setting it to zero. 

\begin{figure}[h]
    \centering
    \captionsetup{justification=raggedright,singlelinecheck=false}
	\includegraphics[width=1.0\textwidth]{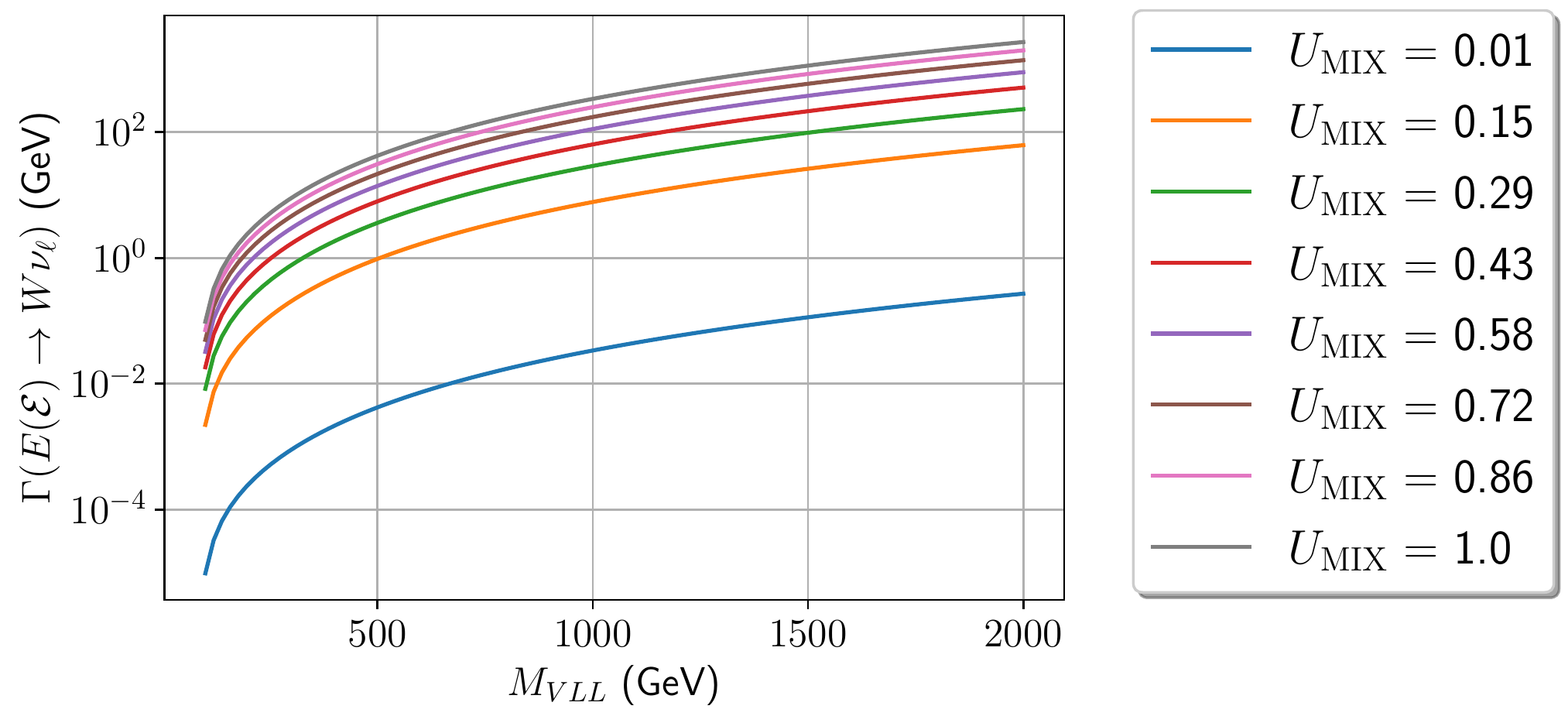}
	\caption{Decay width as a function of the VLL's mass, both displayed in GeV. Each coloured line is representative of a different mixing element, whose numerical value is displayed in the box to the right of the plot.}
	\label{Widths_mass_coupls}
\end{figure}

Another important element to take into consideration is the VLL decay width. In particular, it can be expressed as 
\begin{equation}\label{eq:decay_width}
\begin{aligned}
\Gamma(E(\mathcal{E})\rightarrow W\nu_\ell) = \frac{g^2 U_{\mathrm{MIX}}^2 m_{E,\mathcal{E}}^2}{64\pi M_{W}^2}\qty(1 - \frac{M_W^2}{m_{E,\mathcal{E}}^2})^2\qty(1 + 2\frac{M_W^2}{m_{E,\mathcal{E}}^2}).
\end{aligned}
\end{equation}
Here, $U_{\mathrm{MIX}}^2$ is a non-trivial combination of the neutrino and charged lepton mixing matrices whose specific structure can be determined from the Feynman rules in \cref{eq:Feynman_Rules_Wplus,eq:Feynman_Rules_Wminus,eq:Feynman_Rules_Wplus_1,eq:Feynman_Rules_Wminus_2}. One notes that such a decay width grows as a power law in $m_{E}$, becoming rather significant for large masses as shown in \cref{Widths_mass_coupls}. For example, taking the case where $U_{\mathrm{MIX}} = 1$, one can observe that for a mass of $2~\mathrm{TeV}$, the decay width already surpasses the mass itself with $\Gamma = 2705.4$ GeV. Such an effect can be mitigated for smaller mixing angles as demonstrated in \cref{Widths_mass_coupls}. Indeed, a decay width greater than its mass leads to conceptual issues with respect to the interpretation of such large-width fields as particles. In fact, within the context of Quantum Field Theory, a large value for the width implies that the particle is highly non-local and the narrow-width approximation is no longer valid calling for a more robust framework to describe it (see e.g. \cite{Kuksa:2011ms,Kuksa:2005gq,Kuksa:2009nk}). Such a scenario is not relevant for the charged VLL as the mixing element $U_{\mathrm{MIX}}$ must be small due to EW constraints involving Higgs and Z boson decays. On the other hand, the VLL neutrino partners that reside in the same doublet representation can also decay via a $W$ boson and a charged lepton. In the limit of $m_{\nu_L^\prime} \gg m_\ell$, the expression for the decay width is identical to that of the VLL (with the appropriate changes),
\begin{equation}\label{eq:decay_width_MAJ}
\begin{aligned}
\Gamma(\nu_L^\prime \rightarrow W \ell) = \frac{g^2 U_{\mathrm{MIX}}^2 m_{\nu_L^\prime}^2}{64\pi M_{W}^2}\qty(1 - \frac{M_W^2}{m_{\nu_L^\prime}^2})^2\qty(1 + 2\frac{M_W^2}{m_{\nu_L^\prime}^2}),
\end{aligned}
\end{equation}
where $U_{\mathrm{MIX}}$ can become sizable. Indeed, taking $\ell = \mu$, then, in accordance with the Feynman rules, the mixing elements coming from the left-chiral couplings are given as $\cos{\alpha} \left[U_\nu^\mathrm{doublet}\right]_{55}$. Both $\cos(\alpha)$ and $\left[U_\nu^\mathrm{doublet}\right]_{55}$ are of order $\mathcal{O}(1)$ and therefore the $\nu_L^\prime \rightarrow W \ell$ decay width becomes relevant for large masses. However, notice that the neutrino mass is not independent of its charged VLL counterpart and thus cannot be made arbitrarily small for the width not becoming too large. At leading order, both the VLL and the heavy neutrino masses are controlled by the $M_E$ parameter in \cref{eq:Doubet_case} making them almost degenerate with a small deviation expected to be induced at one-loop level \cite{Thomas:1998wy}. Typically, such a mass splitting is of the order of a few hundred MeV \cite{Thomas:1998wy}, which must not play a sizeable role. With this in mind, and without loss of generality for this article's analysis, we consider both the neutrino and its corresponding charged VLL mass degenerate, ensuring that $\Gamma(\nu_L^\prime \rightarrow W \ell)$ is never unacceptably large. Although the widths must be kept under control, a sizeable value has interesting experimental implications. For instance, it results in wider distributions of kinematic variables such as the mass and the transverse momentum ($p_T$), where large tails can extend into phase space regions not populated by the SM background events.

\section{VLLs: collider phenomenology and flavour constraints}\label{sec:collider_and_g2}

\subsection{Flavour constraints}\label{subsec:Muon_g2}

It follows from interactions between the VLL and the muon that non-zero mixing elements  can trigger Lepton Flavour Violation (LFV) interactions, in particular, those that relate to muon and tau decays. Therefore it is important to confront all points considered in our numerical analysis with the following LFV branching ratios (BR): BR($\mu \rightarrow e\gamma$), BR($\tau \rightarrow e\gamma$), BR($\tau \rightarrow \mu\gamma$), BR($\mu^- \rightarrow e^-e^+e^-$), BR($\tau^- \rightarrow e^-e^+e^-$), BR($\tau^- \rightarrow \mu^-\mu^+\mu^-$), BR($\tau^- \rightarrow e^-\mu^+\mu^-$), BR($\tau^- \rightarrow \mu^-e^+e^-$), BR($\tau^- \rightarrow \mu^-e^+\mu^-$), BR($\tau^- \rightarrow e^-\mu^+e^-$), BR($\tau^+ \rightarrow \pi^0e^+$) and BR($\tau^+ \rightarrow \pi^0\mu^+$). For such a purpose we use the latest version of \texttt{SPheno} \cite{Porod:2011nf} to generate the Wilson coefficient cards, which are then passed to \texttt{flavio} \cite{Straub:2018kue} in order to compute the corresponding LFV observables. It is important to mention that \texttt{SPheno} also computes the BRs i.e.~\texttt{flavio} merely works as an extra layer of added scrutiny, even if both programs' numerical outputs (for the BRs) are well within each other's error and are therefore compatible. We use the 90\% CL experimental limits on the considered LFV observables as reported in the most recent issue of the PDG review \cite{Zyla:2020zbs}.

Last but not least, let us comment that VLLs have long been proposed as a potential explanation for the anomalous magnetic moment of the muon. However, it follows from the smallness of the $\alpha$ mixing angle that neither the singlet nor doublet models discussed in this article can successfully accommodate such an anomaly as couplings of muons to VLLs and to Higgs or Z bosons are too small. This was numerically verified and consistent with the results obtained e.g.~in \cite{Freitas:2014pua}. Therefore, it is not sufficient to extended the SM with one generation of VLLs to explain the muon $g-2$ anomaly and additional new physics with less constrained couplings is necessary. A possibility is to extend the scalar sector as proposed by various authors \cite{Dermisek:2021ajd,Kannike:2011ng,Jegerlehner:2009ry,Dermisek:2013gta,Crivellin:2018qmi,Bonilla:2021ize}.

\subsection{Collider analysis: LHC and muon colliders}\label{subsec:collider_methods}

\begin{figure}[ht!]
    \centering
    \captionsetup{justification=raggedright,singlelinecheck=false}
    \includegraphics[width=0.4\textwidth]{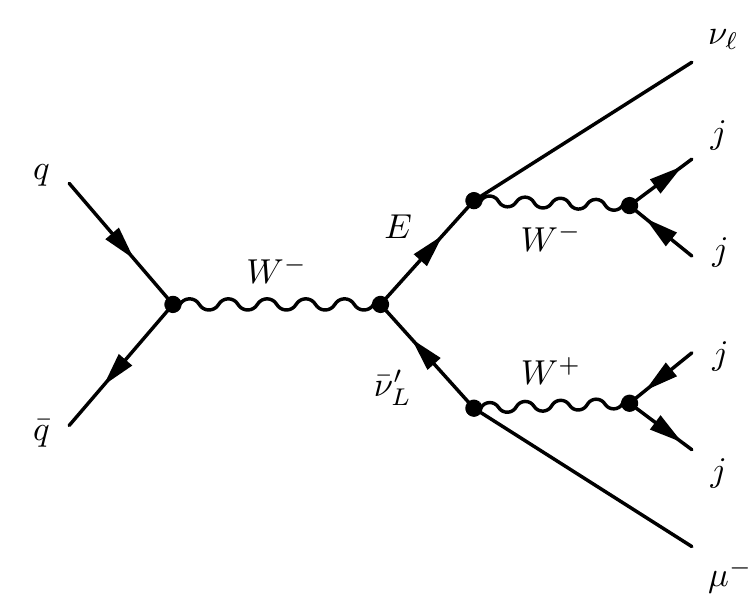}
    \caption{Leading-order Feynman diagram for single production of the doublet VLL. Here, $q$ and $\bar{q}$ correspond to quarks originating from the colliding protons, $E$ represents the VLL and $\nu_{L}^\prime$ denotes the VLL doublet partner. Both $W's$ in the hadronic channel with $j$ indicating first and second generation chiral quarks.}
    \label{fig:Doublet-events}
\end{figure}

The analysis techniques and Deep Learning algorithms used in the current study for both the doublet and the singlet models were first implemented in \cite{Freitas:2020ttd} and \cite{Bonilla:2021ize}, respectively. Starting with the doublet case, the channel we are probing is characterised by a final state with 4 light jets, originating from the decays of two $W$ bosons, a charged lepton, which we take to be the muon, and transverse missing energy (MET), whose origin results from the undetected neutrinos in the final state. At Leading Order (LO), the production diagram, at the LHC, can be seen in \cref{fig:Doublet-events}, where the heavy $\nu_L^\prime$ neutrino belongs to a VLL doublet as shown in \cref{eq:doublets}. For this signal topology the main irreducible backgrounds are the double top-quark production, $t\bar{t}$, $W+$jets (we consider up to 4 jets) and Diboson plus jets. Notice that for the $t\bar{t}$ channel we consider that one of the $W$'s decays in its hadronic channel with the other one decaying into a muon and a neutrino. All backgrounds are generated at LO precision. For the singlet scenario, the final state topology is characterized by one isolated charged lepton from the $W$ boson decay, and transverse missing energy associated with the undetected neutrinos. In particular, we focus on the channel where the charged lepton is a muon. For this signal topology, we consider the irreducible background together with all physics processes that lead to the production of, at least, one muon and up to two jets in the final state, $pp\rightarrow \mu^- \bar{\nu}_\mu$ (+jets)\footnote{In more practical terms, this background is simulated in \texttt{MadGraph} via the commands \texttt{generate p p > mu+ vm \&\& add process p p > mu+ vm j \&\& add process p p > mu+ vm j j} .}.

\begin{figure}[ht!]
    \centering
    \captionsetup{justification=raggedright,singlelinecheck=false}
    \includegraphics[width=0.4\textwidth]{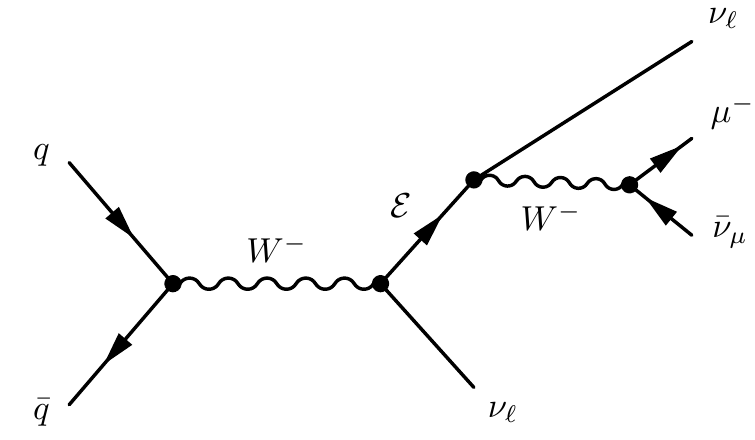}
    \caption{Leading-order Feynman diagram for VLL single production. $q$ and $\bar{q}$ corresponds to quarks originating from the initial protons and $\nu_\ell$ are the SM neutrinos.}
    \label{fig:VLBSM-events}
\end{figure}

For the case of the VLL singlet only one BSM coupling contributes to the decay chain, in particular, the one containing a muon-specific neutrino, whereas for the doublet scenario the couplings involving both the charged and neutral components of the VLL doublet are present. Taking into account the Feynman rules shown in appendix \ref{app:Feyn_rules}, the strength of the relevant left and right chiral charged current projections read as
\begin{equation}
\begin{aligned}
        &g_L^{E \nu_\mu W} = \dfrac{g}{\sqrt{2}} \sin{\alpha} \left[U_\nu^\mathrm{doublet}\right]_{22}, \qquad \qquad
        &&g_L^{\mathcal{E} \nu_\mu W} = \dfrac{g}{\sqrt{2}} \sin{\alpha} \left[U_\nu^\mathrm{singlet}\right]_{22},
        \qquad \qquad \\
        &g_R^{\nu_L^\prime \mu^- W^+} = \dfrac{g}{\sqrt{2}} \left[U_\nu^\mathrm{doublet}\right]_{55}\left[U_R^e\right]_{24},
        \qquad \qquad
        &&g_L^{\nu_L^\prime \mu^- W^+} = \dfrac{g}{\sqrt{2}} \sin{\alpha}\left[U_\nu^\mathrm{doublet}\right]_{55},
        \\
        &g_L^{E \nu_L^\prime W} = \dfrac{g}{\sqrt{2}} \cos{\alpha} \left[U_\nu^\mathrm{doublet}\right]_{55},
        \qquad \qquad
        &&g_R^{E \nu_L^\prime W} = -\dfrac{g}{\sqrt{2}} \left[U^e_R\right]_{24} \left[U_\nu^\mathrm{doublet}\right]_{55}\,.\\
\label{eq:EnuW}
\end{aligned}
\end{equation}
This in turn means that the cross section for the doublet case will be much larger than the one for the singlet case, because the strength of the dominant contribution for the former, in particular the coupling between the VLL and its neutrino counterpart, is proportional to $\cos{\alpha}$ while for the latter all couplings are controlled by $\sin{\alpha}$.  Notice that the single production channel in \cref{fig:VLBSM-events} is valid for both the doublet and singlet scenarios where, in the case of equal angles $g_L^{E \nu_\mu W} = g_L^{\mathcal{E} \nu_\mu W}$. Therefore, it is sufficient to study the singlet model and extract identical conclusions for the doublet case.

The analysis follows a well-defined guideline. Both models are first implemented in \texttt{SARAH} \cite{Staub:2013tta} to generate all interaction vertices as well as all relevant files that interface with Monte-Carlo generators, namely, \texttt{MadGraph5} \cite{Alwall:2014hca} for quark-level matrix-element calculations and \texttt{Pythia8} \cite{Sjostrand:2014zea} for hadronization and showering leading to final state particles. In \texttt{MadGraph5} we simulate proton-proton collisions at a centre-of-mass energy $\sqrt{s} = 14$ TeV, for a total of 250k events for the signal and backgrounds. We also employ the default LO parton-distribution function NNPDF2.3~\cite{Ball:2013hta} which fixes the evolution of the strong coupling constant, $\alpha_s$. Fast-simulation of the ATLAS detector is conducted, with \texttt{Delphes} \cite{deFavereau:2013fsa}. All kinematic distributions are extracted with the help of \texttt{ROOT} \cite{Brun:1997pa}. At this stage, we also impose selection criteria, to maximise the signal significance. In particular, all events must satisfy
\begin{equation}\nonumber
\begin{aligned}
&p_T(\mu^-) > 25 \text{ GeV}, \\
&\rm{MET}  > 15 \text{ GeV} \,,  \\
&\abs{\eta(\mu^-)} \leq 2.5 \, \rm{and}\\
&\abs{\eta(j)} \leq 2.5\, .\\
\end{aligned}
\end{equation}
\\
where $\eta$ represents the muon pseudo-rapidity\footnote{The pseudorapidity of a particle is defined as $\eta = -\ln(\tan(\theta/2))$, where $\theta$ is its polar angle.}. Additionally, for the doublet topology, which involves jets in the final state, we require them to be tagged as originating from light chiral quarks, \textit{i.e.}, they cannot be tagged as b-jets. Note that the signal production represented in Fig.~\ref{fig:VLBSM-events}, is characterized by having no jets in the final states. While it is true that the lack of no jet candidates at any given event in a hadron collider is perhaps too unrealistic, the absence of such selection can also cause complications in the Deep Learning algorithms used later in the analysis, since signal/background classes are incredibly unbalanced and may lead to over-fitting problems. 
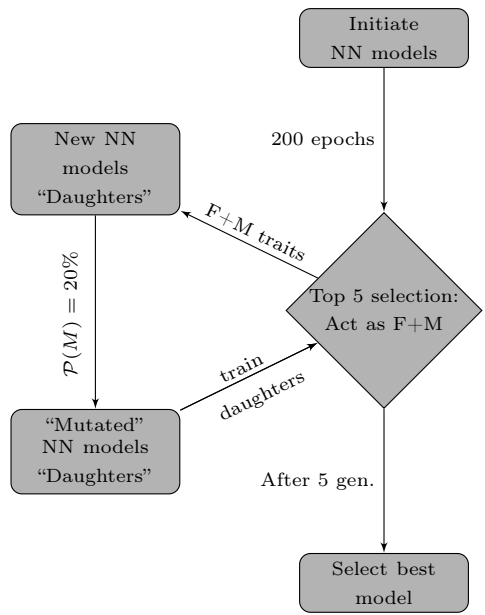
\begin{figure}[th!]
	\centering
	\tikzstyle{Rectangulo} = [draw, rectangle, fill=black!30, text width=6em, text centered, minimum height=2em]
	\tikzstyle{Diamante} = [draw, diamond, fill=black!30, text width=6em, text badly centered, inner sep=0pt]
	\tikzstyle{Linha} = [draw, -latex']
	\resizebox{0.35\textwidth}{!}{\begin{tikzpicture}[node distance = 1.5cm, auto]
		\node [Rectangulo, rounded corners] (step1) {\scriptsize Initiate NN models};
		\node [Diamante, below of=step1,node distance=3.5cm] (step2) {\scriptsize Top 5 selection: Act as F+M};
		\node [Rectangulo, rounded corners, below of=step2, node distance=3.5cm] (step3) {\scriptsize Select best model};
		\node [Rectangulo, rounded corners, above left of=step2, node distance=2.0cm, above left=0.4cm, left=0.8cm] (step4) {\scriptsize New NN models ``Daughters''};
		\node [Rectangulo, rounded corners, below left of=step2, node distance=2.0cm, below left=0.4cm, left=0.8cm] (step5) {\scriptsize ``Mutated'' NN models \\ ``Daughters''};
		\path [Linha] (step1) -- node [left] {\scriptsize 200 epochs} (step2);
		\path [Linha] (step2) -- node [left] {\scriptsize After 5 gen.} (step3);
		\path [Linha] (step2) -- node [above=0.50cm, right=-0.65cm, rotate=-26] {\scriptsize F+M traits} (step4);
		\path [Linha] (step4) -- node [left, rotate=90, below=-0.3cm, left=-0.9cm] {\scriptsize $\mathcal{P}(M) = 20\%$}(step5);
		\path [Linha] (step5) node [above=0.90cm , right=1.50cm, rotate=26] {\scriptsize train} -- (step2);
		\path [Linha] (step5) node [below=-0.40cm, right=1.50cm, rotate=26] {\scriptsize daughters} -- (step2);
		\end{tikzpicture}}
	\caption[]{Flowchart representative of all iterative steps involved in the genetic algorithm that we employ in this work.}
	\label{fig:EVO-algo}
\end{figure}
The kinematics of final states for both signal and background events are translated into tabular datasets and used as inputs for neural models, whose job is to separate the signal from the background. The neural network is constructed using \texttt{Keras} \cite{chollet2015keras}. For classification of the singlet model, we choose a set of 5 observables computed in the laboratory frame. Low-level features include the muon observables $\cos(\theta_{\mu^-})$, $\eta(\mu^-)$, $\phi(\mu^-)$ and $p_T(\mu^-)$ as well as the MET. 

For the doublet model, a richer final state is present and, as such, one can compute a much more complete list of observables. Do note that, for this physics scenario, the final states contain 4 light-jets, which implies that one cannot distinguish between the jets originating from $W^+$ and those originating from $W^-$. Therefore, for reconstructed observables, we consider all possible combinations of jets that can be used. These are $C$ = $(j_1,j_2)$, $(j_1,j_3)$, $(j_1,j_4)$, $(j_2,j_3)$, $(j_2,j_4)$ and $(j_3,j_4)$, where we define $j_1$ as the leading jet, with the highest $p_T$, and $j_4$ as the subleading jet, with the lowest $p_T$. The full list of observables that are used in the training are shown in \cref{tab:vars_doublet}.

\begin{table}[H]    
	\centering
	\captionsetup{justification=raggedright,singlelinecheck=false}
	\resizebox{1.0\textwidth}{!}{\begin{tabular}{|c|c|c|c|}
			\toprule
			\hline
			& Dimension-full & \multicolumn{2}{|c|}{Dimensionless} \\
			\hline
			\hline
			\makecell{Doublet VLL}  & \makecell{$p_T(\mu^-)$, $p_T(j_1)$, $p_T(j_2)$, $p_T(j_3)$ MET, \\
				$p_T(j_4)$, $E(\mu^-)$, $E(j_1)$, $E(j_2)$, \\
				$E(j_3)$, $E(j_4)$, $M(j_1,j_2)$, $M(j_1,j_3)$, \\
				$M(j_a,j_b)$, $p_T(j_a,j_b)$, $M(j_a,j_b,\mu^-)$,\\ $p_T(j_a,j_b,\mu^-)$, $M(j_a,j_b,\mathrm{MET})$, $p_T(j_a,j_b,\mathrm{MET})$\\} & 
			\makecell{
				$\cos(\theta_{j_a j_b})$, $\cos(\theta_{j_a \mu^-})$, \\
                 $\eta(j_1)$, $\eta(j_2)$, $\eta(j_3)$, $\eta(j_4)$, \\ $\eta(\mu^-)$, $\phi(j_1)$, $\phi(j_2)$, $\phi(j_3)$, \\ $\eta(j_4)$, $\eta(j_a,j_b)$, $\phi(j_a,j_b)$, \\
                 $\eta(j_a,j_b,\mu^-)$, $\eta(j_a,j_b,\mu^-)$} & 
			\makecell{$\Delta R(j_a, j_b)$, $\Delta R(j_a, \mu^-)$, $\Delta \phi(j_a, j_a)$, \\
				$\Delta \phi(j_a, \mu^-)$, $\Delta \theta (j_a, j_b)$, $\Delta \theta (j_a, \mu^-)$.} \\
			\hline
			\hline
	\end{tabular}}
	\caption{Angular and kinematic distributions for the analysis of the doublet production topology. All observables are computed in the laboratory frame of reference. To simplify notation, we define $(j_a,j_b)$ as corresponding to all jet combinations, as described in the text. Variables that involve combinations of final states, e.g. $p_T(j_a,j_a,\mu^-)$, are reconstructed from the states indicated inside the parenthesis.}
	\label{tab:vars_doublet}
\end{table}

To optimize the neural network architecture, we employ a genetic algorithm following the same steps as described in \cite{Freitas:2020ttd,Bonilla:2021ize} and schematically represented in the diagram of \cref{fig:EVO-algo}. The algorithm begins by first generating an arbitrary number of neural networks, whose architecture is determined by randomly pooling a list of predefined hyperparameters (number of layers, activation functions, regularisers, etc.). We then train each individual network over the data for a given number of epochs. From the trained networks, we pick the top five best performing ones. From these, we create Father-Mother pairs, where we combine 50\% of the Father's traits (that is, its hyperparameters) and 50\% of the Mother traits to construct new neural architectures, which we dub as Daughters. We also consider the possibility of mutation, that is, after the Daughter networks have been built, we consider that the hyperparameters may change to another, with the probability of $\mathcal{P}(M) = 20\%$. We then train the Daughter networks for a given number of epochs and repeat the procedure for a set of generations. Finally, we select the best performing network based on some metric of choice. In our analysis, the algorithm is designed to maximise the Asimov statistical significance, based on a earlier work of Elwood and Adam in \cite{Elwood:2018qsr}. The Asimov metric is defined as \cite{Cowan:2010js}
\begin{equation}\label{eq:Asimov_sig}
\mathcal{Z_A} = \Bigg[2\Bigg((s + b)\ln\Bigg(\frac{(s+b)(b+\sigma_b^2)}{b^2 + (s+b)\sigma_b^2}\Bigg) -\frac{b^2}{\sigma_b^2}\ln\Bigg(1+\frac{\sigma_b^2 s}{b(b+\sigma_b)}\Bigg)\Bigg)\Bigg]^{1/2},
\end{equation}
where $s$ is the number of signal events, $b$ is the number of background events and $\sigma_b$ is the uncertainty of the background. As part of the optimization procedure, we consider the same list of hyperparameters as in our previous work \cite{Freitas:2020ttd}:
\begin{itemize}
    \item number of hidden layers: 1 to 5
    \item number of neurons in each layer: 256, 512, 1024 or 2048
    \item kernel initializer: 'normal','he normal','he uniform'
    \item L2 regularization penalty: 1e-3, 1e-5, 1e-7
    \item activation function: 'relu', 'elu', 'tanh', 'sigmoid'
    \item optimizer: 'adam', 'sgd', 'adamax', 'nadam'
\end{itemize}
\noindent Our evolutionary algorithm is initialized by building a set of ten NNs. The parameters are chosen randomly from the previous lists. Each network is trained up 200 epochs and if any improvement is not observed by at least 5 epochs, then the training stops. Note that a subset of the neural network properties are not subject to optimisation but instead remain fixed during the runs. Namely, we consider the following
\begin{itemize}
	\item The input data of the NNs are standardised, that is, input vectors have a mean of zero and a standard deviation of 1. All observables were extracted from the \texttt{ROOT} files and outputed into dataframes. The data is reshuffled and then divided into training samples (80\% of the total data) and validation samples (the remainder 20\%). We also consider cross-validation  with a five-fold scheme during training. The statistical significance is computed based on the trained NN predictions of the validation data.
	\item We employ a cyclic learning rate during the training phase with 0.01 initial value and maximal value of 0.1.
    \item The output of the NN is a vector of probabilities such that we define a signal for an output with probability greater than 0.5, otherwise it is considered as a background.
	\item Batch size of 32768. 
	\item The best NN is selected based on the Asimov metric. The loss function for this case is defined as the inverse of Eq.~\eqref{eq:Asimov_sig}, such that when the loss function is minimised, the Asimov significance is maximised. We also consider a fixed value of the background uncertainty to $\sigma_b = 10^{-1}$.
	\end{itemize}

An additional consideration is the fact that our datasets are unbalanced, which is in part a result of the selection criteria imposed on the final states, reducing the allowed phase space for both signal and backgrounds. Unbalanced datasets can lead to overfitted networks and, as such, must be properly dealt with. In this work, we have utilised the Synthetic Minority Oversampling Technique (SMOTE) \cite{2011arXiv1106.1813C}, which oversamples the minority classes of our training data. Do note that this algorithm is only employed to the training dataset and we do not perform any re-sampling of the validation samples.

With the LHC being a proton collider, production of coloured particles by the strong interaction is heavily favoured, compared to identical processes involving electroweak bosons. As VLLs are colour singlets, they can only be produced via this last interaction at LO. It is then worth mentioning that, in addition to the obvious motivation of studying VLL production at the LHC already during RUN3 and, later on, in its high luminosity phase (HL-LHC), it is of utmost importance to understand the sensitivity with which these new particles can be probed, at future colliders. 

\begin{figure}[ht!]
    \centering
    \captionsetup{justification=raggedright,singlelinecheck=false}
    \includegraphics[width=0.6\textwidth]{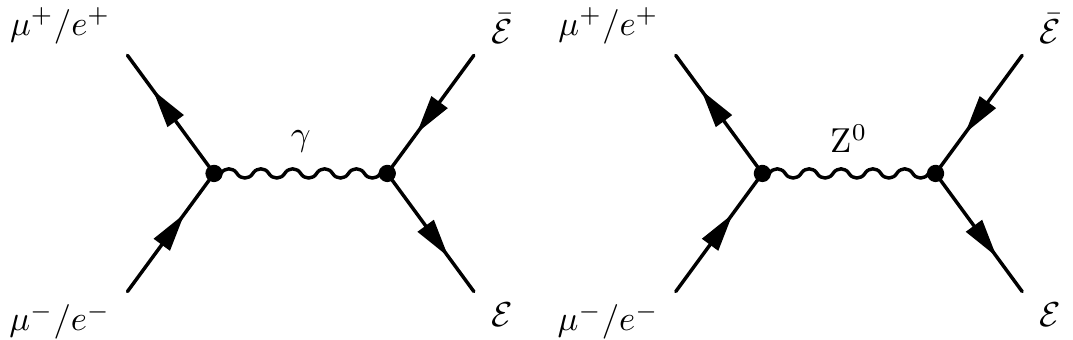}
    \caption{LO Feynman diagrams for VLL double production at a lepton/anti-lepton collider via exchange of either a photon or a $\mathrm{Z^0}$ boson.}
    \label{fig:schannel-muon}
\end{figure}

In particular, there has been an active discussion within the community about $e^+e^-$ colliders like the Compact Linear Collider (CLIC) \cite{Charles:2018vfv,CLIC:2016zwp} or the International Linear Collider (ILC)~\cite{Baer:2013cma}, and more recently, on the possibility of building a $\mu^-\mu^+$ collider~\cite{Muon_colliders_1,Muon_colliders_2}. 
Besides offering cleaner environments, when compared to hadronic machines, production via electroweak processes is favoured, hence, VLLs should have a higher chance of being observed, in case they exist, in this type of machines. For this purpose, we perform numerical and analytical computations with \texttt{FeynCalc} \cite{Shtabovenko:2020gxv} for the pair-production of VLLs whose tree-level diagrams can be seen in Fig.~\ref{fig:schannel-muon}. As previously stated, singlets have lower couplings compared to doublets, hence we discuss the prospects for VLL discovery at lepton colliders for the singlet scenario and not the doublet, as this can be seen as the worst-case scenario. Do note, however, that the conclusions one takes from the singlet model can be easily generalized to the doublet model. That is, the only difference between the collider studies for the two scenarios are the values of the couplings. 

In conclusion, the single VLL production is favoured at the LHC for the doublet case, due mainly to the strength of the couplings. Double production of VLLs has the obvious disadvantage of the need to produce two heavy states and a more elaborate final state to detect. Additionally, as it was shown in previous works \cite{Freitas:2020ttd,Bonilla:2021ize}, double production topologies are sub-leading when compared with single production ones and, as such, are not considered in this work.
If one wants to probe the singlet model, and assuming that it cannot be done at the LHC, the future lepton colliders, where single production is precluded at LO, may give us enough VLLs in the double production channel to test the singlet VLL scenario.

\section{Results}\label{sec:Results}

In this section, we present a numerical analysis focusing on the LHC and future lepton colliders. In \cref{mixings_doublet,mixings_singlet} we show two distinct viable scenarios (one for the doublet and one for the singlet model), where all couplings are fixed and only the VLL mass varies. The colour coding, black for the doublet model and green for the singlet case, merely serves to distinguish each of the benchmark scenarios and identify them throughout the text whenever necessary. We first consider a benchmark point where we set the VLL mass to $M_{\rm VLL}=700$~GeV.
The remaining BSM neutrinos present in the doublet model share, at LO, the same mass with their $\SU{2}{L}$ doublet VLL counterpart and, as noted in \cref{eq:EnuW}, these neutrinos efficiently decay into muons and $W^\pm$ bosons, with an interaction strength of the order of the weak gauge coupling,
where, for the considered benchmarks, $\left[ U^\mathrm{doublet}_\nu \right]_{55} \approx \left[ U_R^e \right]_{24} \approx 1$. Indeed, the right chiral component of the coupling is the dominating factor, as the left-coupling is suppressed by a factor of $\sin(\alpha)$, which is of $\mathcal{O}(10^{-2})$ in our numerical analysis. The selected benchmark point, represented in black, for the doublet model, compatible with all flavour observables discussed above reads as
\begin{equation}\label{mixings_doublet}
\begin{aligned}
&U^e_L = \begin{bmatrix}
1 & 0 & 0 & 0 \\
0 & 0.9997545 & 0 & -0.02215850 \\
0 & 0 & 1 & 0 \\
0 & -0.02215850 & 0 & 0.9997545
\end{bmatrix}, \quad
U^e_R = \begin{bmatrix}
1 & 0 & 0 & 0 \\
0 & 0.999842 & 0 & 0.0177691 \\
0 & 0 & -1 & 0 \\
0 & 0.0177691 & 0 & 0.999842
\end{bmatrix}, \\[0.5em]
&U_\nu = \begin{bmatrix}
U_{\mathrm{PMNS}} & 0 \\
0 & U_\nu^{\mathrm{BSM}}
\end{bmatrix}, \quad U_\nu^{\mathrm{BSM}} = \begin{bmatrix}
1 & 0 & 0 \\
0 & -0.960772 & -0.277338 \\
0 & 0.277338 & -0.960772
\end{bmatrix},
\end{aligned}
\end{equation}
whereas for the singlet scenario, in green, one has
\begin{equation}\label{mixings_singlet}
\begin{aligned}
&{\green U^e_L = \begin{bmatrix}
1 & 0 & 0 & 0 \\
0 & 0.9995324 & 0 & -0.03057725 \\
0 & 0 & 1 & 0 \\
0 & -0.03057725 & 0 & 0.9995324
\end{bmatrix}}, \quad
{\green U^e_R = \begin{bmatrix}
1 & 0 & 0 & 0 \\
0 & 1 & 0 & -1.57282\times 10^{-5} \\
0 & 0 & -1 & 0 \\
0 & -1.57282 \times 10^{-5} & 0 & -1
\end{bmatrix}}, \\[0.5em]
&U_\nu = \begin{bmatrix}
U_{\mathrm{PMNS}} & 0 \\
0 & 1
\end{bmatrix}. \\[0.5em]
\end{aligned}
\end{equation}
The neutrino mixing is identical in the two considered cases and therefore is always represented in black.

\begin{figure}[h]
    \centering
    \captionsetup{justification=raggedright,singlelinecheck=false}
	\includegraphics[width=0.80\textwidth]{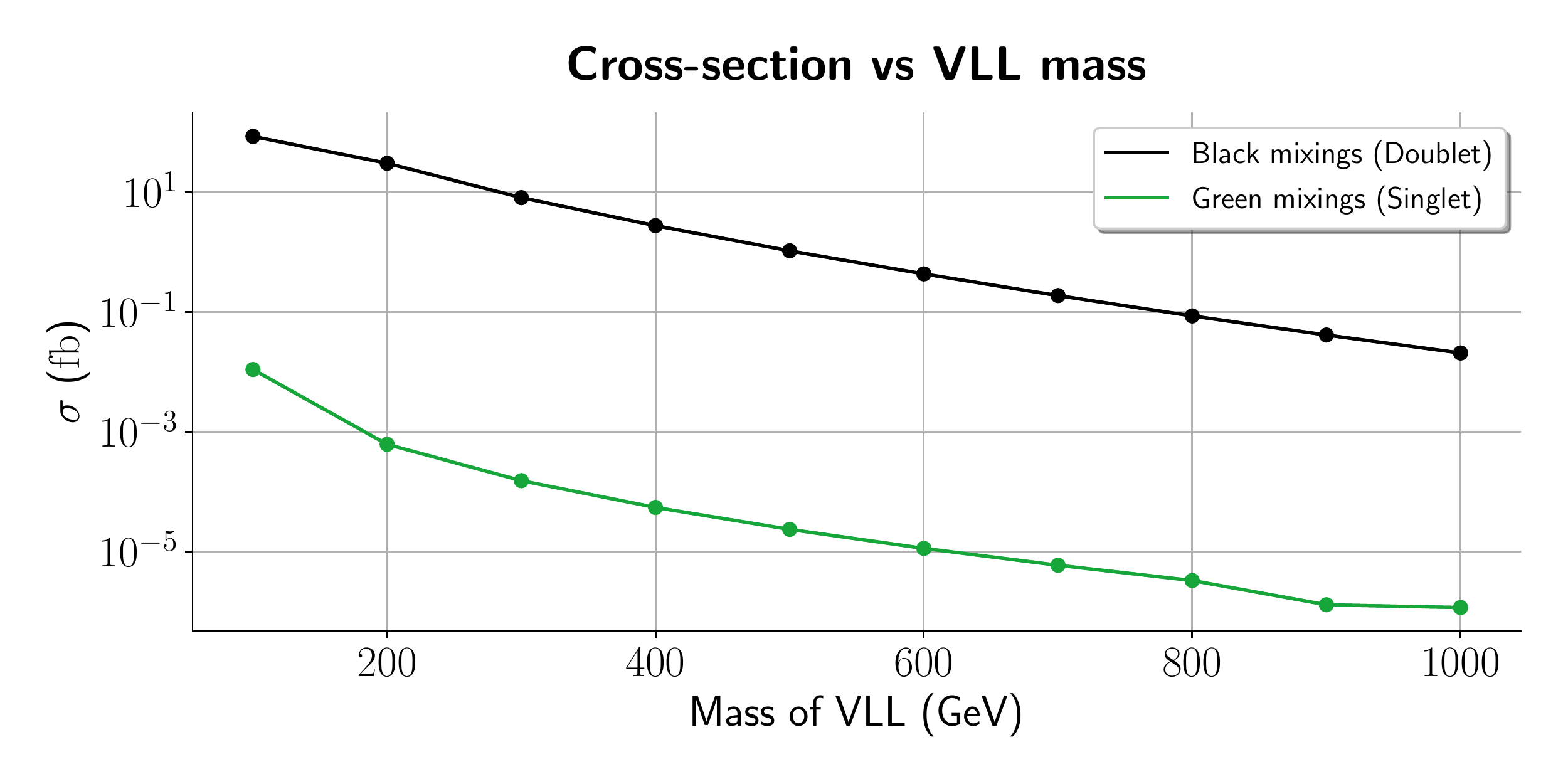}
	\caption{Production cross section for both doublet and singlet topologies, in femtobarn (fb), as a function of the VLL's mass, in GeV. The $y$-axis is shown in logarithmic scale. The black and green curves correspond to the same colour mixing scenarios in \cref{mixings_doublet,mixings_singlet}.}
	\label{fig:xsecs_VLLs_LHC}
\end{figure}

It follows from \cref{eq:EnuW} and the numerical values shown above that, upon applying the selection criteria described in Section~\ref{subsec:collider_methods}, the signal and background cross sections are
\begin{equation}\nonumber
\begin{aligned}
&\sigma(\text{doublet}): 0.19 \hphantom{.} \mathrm{fb},\\
&\sigma(\text{singlet}): 5.94 \times 10^{-6} \hphantom{.} \mathrm{fb},\\
&\sigma(p p \rightarrow \mu^- \bar{\nu}_\mu \text{+ up to 2 jets): } 4.79\times 10^{6} \hphantom{.} \mathrm{fb},\\
&\sigma(t\bar{t}): 13616.16 \hphantom{.} \mathrm{fb},\\
&\sigma(W+\mathrm{jets}): 102448.54 \hphantom{.} \mathrm{fb},\\
&\sigma(\mathrm{Diboson}+{\mathrm{jets}}): 296.94 \hphantom{.} \mathrm{fb},
\end{aligned}
\end{equation}
It is evident that the doublet model cross section is far larger than that of the singlet. This is indeed expected as the dominant contribution in the singlet model contains a $\sin{\alpha}$ suppression factor as one can see in \cref{eq:EnuW}. The dependency of the production cross section with the VLL mass is shown in \cref{fig:xsecs_VLLs_LHC}. 
One can now estimate the total expected number of events, given the cross section above, after event selection. Assuming the target luminosity of the HL-LHC, $\mathcal{L} = 3000$ $\mathrm{fb^{-1}}$ and that the expected number of events is given by $N = \sigma \mathcal{L}$, we have for the doublet case $N = 570.0$ events, while for the singlet we obtain $N = 0.01782$ events. For the singlet model, this implies that the production cross section is not large enough to generate one event at the LHC. For this reason it is meaningless to present the statistical significance for the singlet model for such a heavy VLL. However, we see from \cref{fig:xsecs_VLLs_LHC} that, for lighter singlet VLLs, in particular for their masses of $100$ and $200~\mathrm{GeV}$, one has $N \approx 30$ and $N \approx 3$, respectively. Thus, it becomes possible to produce them at the HL-LHC motivating a further full analysis.

The kinematic features used in the Deep Learning analysis can be seen in \cref{fig:Doublet_kin,fig:Singlet_kin} of \cref{app:Kinematic_dist} for the doublet and singlet models, respectively. For the singlet model, the main variables allowing for a good discrimination between the signal and the background are the transverse momentum of the muon i.e., $p_T(\mu^-)$  as well as the MET distribution. These distributions are characterized by long tails at higher energies where the SM background is no longer present. On the other hand, the angular distributions for the cosine of the polar angle, as well as the azimuthal angle, offer the least discriminating power. This follows from the fact that both signal and background have a similar shape. The pseudo-rapidity distributions can also be used in the discrimination since signal events tend to peak around $\eta \sim 0$ whereas the SM backgrounds spread out over the entire $\abs{\eta} \leq 2.5$ region. In particular, for the $\eta(\mu^-)$ distributions, the SM backgrounds spread uniformly in the entire range. For the doublet model, we note that the kinematic distributions offer the best discriminating power, with both the $p_T$ and mass distributions peaking at higher energies than those of the main irreducible backgrounds. While $\Delta \theta$ distributions closely follow the SM background prediction, the $\Delta\phi$ and $\Delta R$ distributions are distinct from those of the SM, with $\Delta \phi$ possessing a double peak structure near zero, whereas $\Delta R$ have its maximum at zero.

With this information at hand, one can construct multi-dimensional distributions to be used as inputs for a neural network that solves a classification task. This is done via an evolution algorithm to optimise the various hyper-parameters of the neural model, whose metric to be maximised is the Asimov statistical significance. For completeness, we present our results for different statistical models, some more conservative than others. In this article we use the same measures as in \cite{Freitas:2020ttd,Bonilla:2021ize}, which include:
\begin{enumerate}
    \item The Asimov significance, $\mathcal{Z}_A$, with 1\% systematic uncertainty \footnote{{ The dominant backgrounds of the analysis ($t\bar{t}$ and W+jets) have currently their cross sections measured with $\mathcal{O}$(2\%) precision with an important contribution from the integrated luminosity measurement which strongly dominates the total uncertainty. When preparing for the European Strategy for Particle Physics reports that both ATLAS and CMS contributed to, a range of 1-1.5\% was considered as a guideline for the luminosity uncertainty of the HL-LHC \cite{CERN_HLLHC}. Furthermore, in a recent update \cite{ATLAS:2022hro} this recommendation was further improved to $0.83\%$, based on the full run-II accumulated dataset. As such, a global of 1\% in our backgrounds can be seen as a realistic target for both run-III and the HL phase of the LHC.}}, as defined in Eq.~\eqref{eq:Asimov_sig};
    \item A less conservative version of the Asimov significance, which we dub as $\mathcal{Z}(<1\%)$. In this measure, we assume that backgrounds are known with an error of $10^{-3}$. Of all measures, this is the most lenient one and typically offers the most significant results;
    \item The more traditional metric, $s/\sqrt{s+b}$.
\end{enumerate}

\begin{figure*}[]
    \captionsetup{justification=raggedright,singlelinecheck=false}
	\subfloat{{\includegraphics[width=0.35\textwidth]{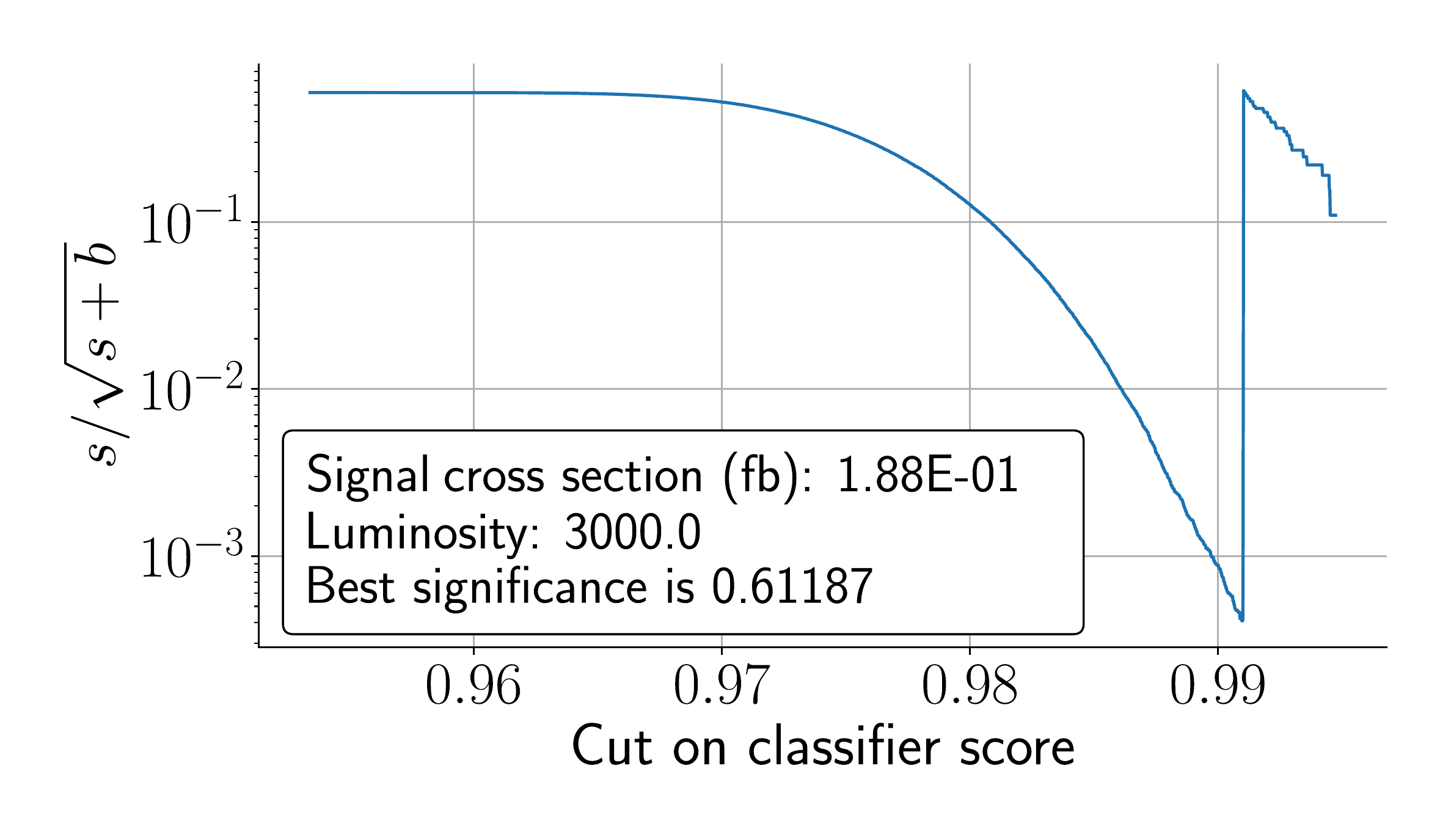} }} 
	\subfloat{{\includegraphics[width=0.35\textwidth]{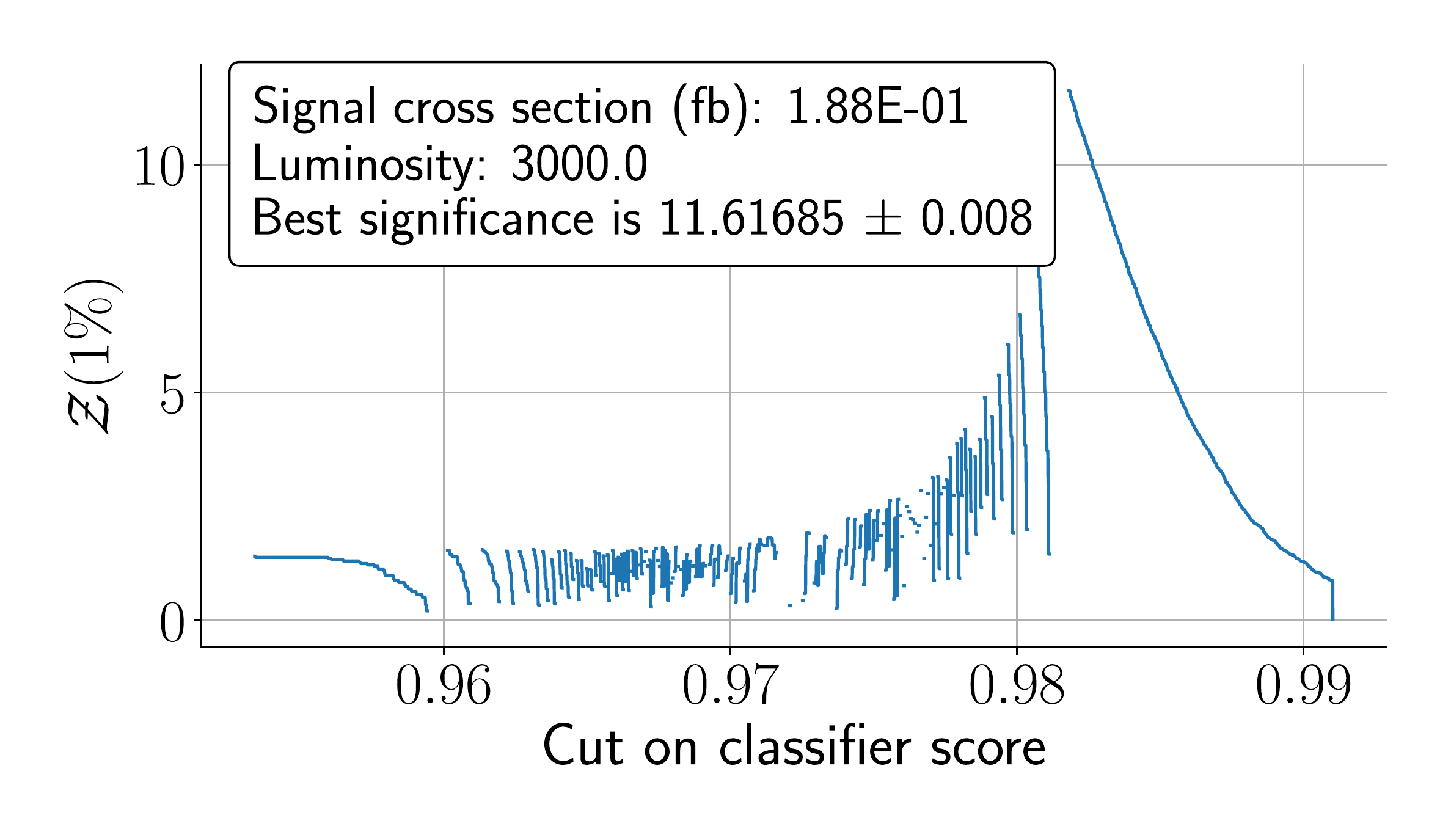} }} 
	\subfloat{{\includegraphics[width=0.30\textwidth]{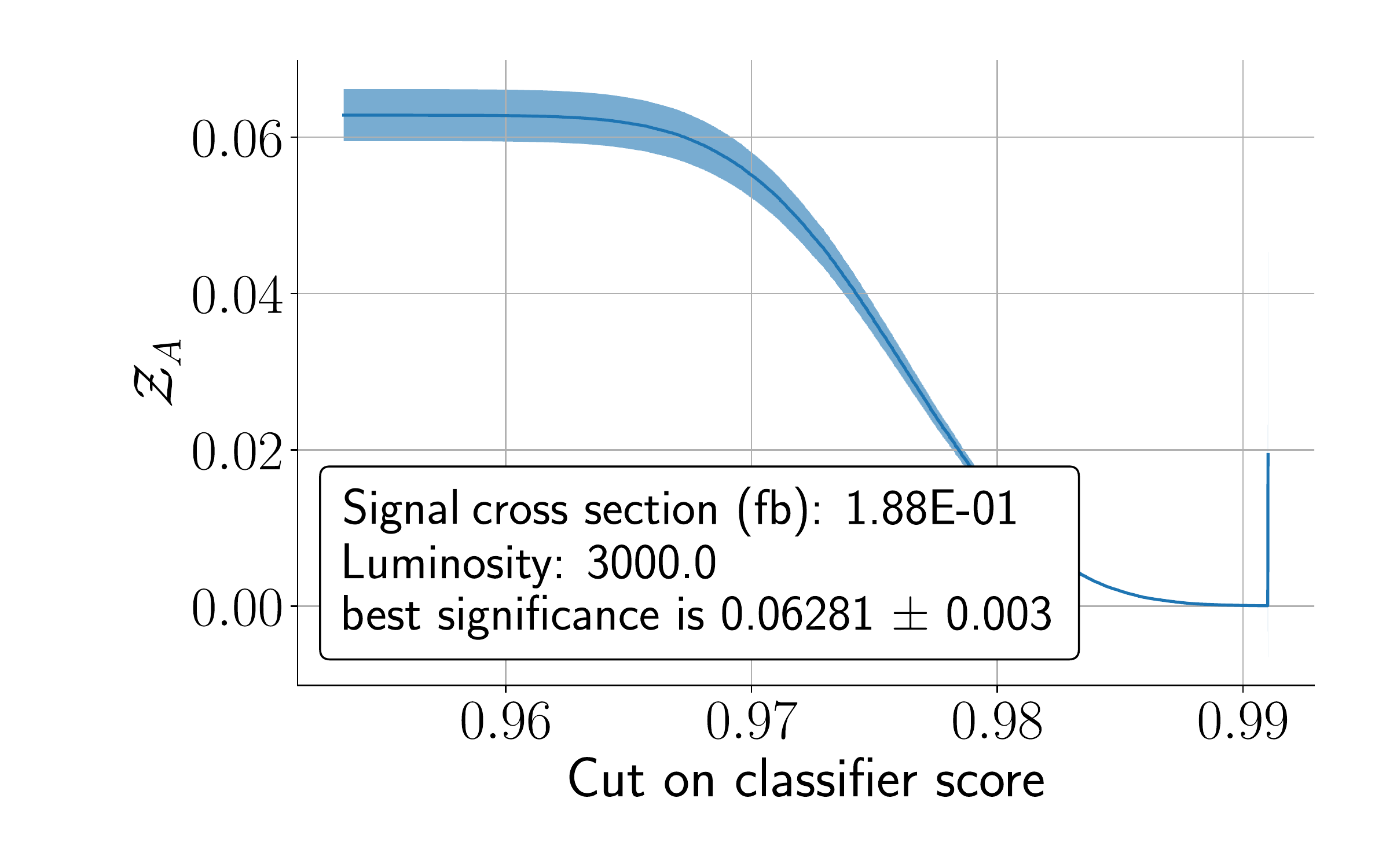} }}\\
	\caption{Statistical significance as a function of the classifier score given by a neural network for different metrics, assuming a doublet-type VLL with $M_{\mathrm{VLL}}=700$ GeV and a collider luminosity $\mathcal{L} = 3000$ $\mathrm{fb}^{-1}$. The computation of the statistical significance is made using the best neural network that the evolution algorithm found. From left to right we plot the significance $s/\sqrt{s+b}$ in (a), the adapted Asimov significance where we assume a background uncertainty of $10^{-3}$ in (b), and the Asimov significance with systematics of 1\% in (c).
		\label{fig:sig_plots}}
\end{figure*}

\begin{figure}[ht!]
    \centering
    \captionsetup{justification=raggedright,singlelinecheck=false}
	\includegraphics[width=0.7\textwidth]{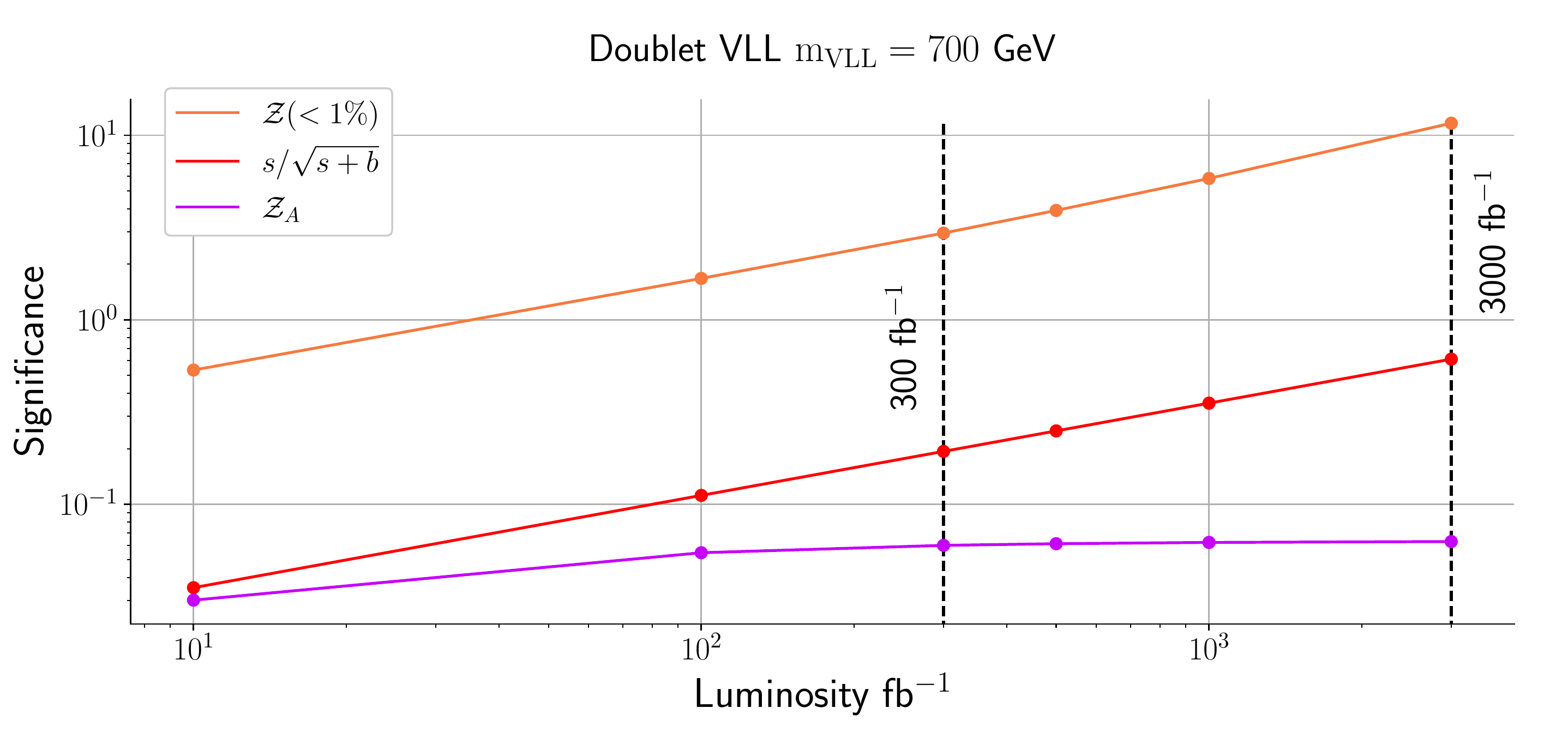}
	\caption{Statistical significance as a function of the collider luminosity, for a fixed doublet VLL mass of $700~\mathrm{GeV}$ (mixing in \cref{mixings_doublet}). Target luminosities of the HL-LHC (3000 $\mathrm{fb}^{-1}$) and Run III (300 $\mathrm{fb}^{-1}$) are marked with vertical dashed black lines. Both axes are shown in logarithmic scale.}
	\label{fig:Sig_dif_statistics}
\end{figure}

With this in mind, we compute the significances, for the doublet scenario, in a wide range of masses, from 100 to 1000~GeV, in steps of 100~GeV. In particular, we plot the various metrics as a function of the neural network score in Fig.~\ref{fig:sig_plots}, taking a VLL mass equal to 700~GeV, for illustration purposes. For this mass point and, for an integrated luminosity of $\mathcal{L} = 3000$ $\mathrm{fb^{-1}}$, we obtain
\begin{equation}\nonumber
\begin{aligned}
& s/\sqrt{s+b} = 0.61\sigma,\\
& \mathcal{Z}(<1\%) = 11.62\sigma,\\
& \mathcal{Z}_A = 0.063\sigma,
\end{aligned}
\end{equation}
 where we note that we can exclude (or claim an hypothetical discovery) for a VLL with such a mass, since the $\mathcal{Z}(<1\%)$ metric offers a statistical significance larger than $5\sigma$. However, one must keep in mind that this metric is the least conservative of the those considered in this work. On the other hand, the Asimov metric is the stringiest and most conservative one resulting in a tiny significance for this particular point. It is also important to study the role of different values of the luminosity. As such, in Fig.~\ref{fig:Sig_dif_statistics}, we show the evolution of the significance as a function of the collider's luminosity for a VLL with a mass of $700~\mathrm{GeV}$. In particular, we highlight with dashed black vertical lines the target luminosities at Run III (300 $\mathrm{fb^{-1}}$) and at the HL-LHC (3000 $\mathrm{fb^{-1}}$). Focusing on Run III, we obtain
\begin{equation}\nonumber
\begin{aligned}
& s/\sqrt{s+b} = 0.19\sigma,\\
& \mathcal{Z}(<1\%) = 2.95\sigma,\\
& \mathcal{Z}_A = 0.059\sigma,
\end{aligned}
\end{equation}
such that, for a $700~\mathrm{GeV}$ VLL, one does not expect any significant excess and therefore can not extract conclusions.

For an inclusive picture we perform a mass scan in the range $M_{\mathrm{VLL}} \in [100, 1000]$ GeV, whose numerical values for the couplings are fixed to those shown in \cref{mixings_doublet}. On the other hand, it follows from the discussion in the previous paragraphs that only two example points are selected for the singlet model, both featuring the green mixing matrices in \cref{mixings_singlet}. We also vary the mass of the heavy left-handed neutrinos, present in the doublet model, such that they share the same mass with their charged $\SU{2}{L}$ doublet counterpart. In \cref{tab:Sig_VLLs} our results for the scan are summarized, where the VLL masses and the calculated significance for both Run III and the HL-LHC upgrade are shown.

\begin{table}[htb!]
	\centering
	\resizebox{0.70\textwidth}{!}{\begin{tabular}{c||ccc||ccc||}
			\multirow{2}{*}{Mass of VLL (GeV)} & \multicolumn{3}{c||}{$300$ $\mathrm{fb^{-1}}$}                                                               & \multicolumn{3}{c||}{$3000$ $\mathrm{fb^{-1}}$}                                      \\ \cline{2-7} 
			& $s/\sqrt{s+b}$                          & $\mathcal{Z}(<1\%)$                         & $\mathcal{Z}_A$                & $s/\sqrt{s+b}$                           & $\mathcal{Z}(<1\%)$                         & $\mathcal{Z}_A$                  \\ \hline
			$100$                      & \multicolumn{1}{c|}{$75.32$} & \multicolumn{1}{c|}{$81.76$} & \multicolumn{1}{c||}{$25.01$}       &\multicolumn{1}{c|}{$238.17$} & \multicolumn{1}{c|}{$258.55$} & \multicolumn{1}{c||}{$26.13$}  \\[0.2em]
			$200$                      & \multicolumn{1}{c|}{$28.97$} & \multicolumn{1}{c|}{$29.92$} & \multicolumn{1}{c||}{$9.37$}       &\multicolumn{1}{c|}{$91.62$} & \multicolumn{1}{c|}{$94.61$} & \multicolumn{1}{c||}{$9.81$}  \\[0.2em]
			$300$                      & \multicolumn{1}{c|}{$8.02$} & \multicolumn{1}{c|}{$8.09$} & \multicolumn{1}{c||}{$2.56$}       &\multicolumn{1}{c|}{$25.35$} & \multicolumn{1}{c|}{$25.58$} & \multicolumn{1}{c||}{$2.68$}  \\[0.2em]
			$400$                      & \multicolumn{1}{c|}{$2.76$} & \multicolumn{1}{c|}{$2.77$} & \multicolumn{1}{c||}{$0.88$}       &\multicolumn{1}{c|}{$8.73$} & \multicolumn{1}{c|}{$8.76$} & \multicolumn{1}{c||}{$0.92$} \\[0.2em]
			$500$                      & \multicolumn{1}{c|}{$1.05$} & \multicolumn{1}{c|}{$2.70$} & \multicolumn{1}{c||}{$0.33$}       &\multicolumn{1}{c|}{$3.32$} & \multicolumn{1}{c|}{$11.19$} & \multicolumn{1}{c||}{$0.35$}   \\[0.2em]
			$600$                      & \multicolumn{1}{c|}{{$0.43$}} & \multicolumn{1}{c|}{{$2.82$}} & \multicolumn{1}{c||}{{$0.14$}}       &\multicolumn{1}{c|}{{$1.37$}} & \multicolumn{1}{c|}{{$11.43$}} & \multicolumn{1}{c||}{{$0.14$}}  \\[0.2em]
			$700$                      & \multicolumn{1}{c|}{$0.19$} & \multicolumn{1}{c|}{$2.95$} & \multicolumn{1}{c||}{$0.060$}       &\multicolumn{1}{c|}{$0.61$} & \multicolumn{1}{c|}{$11.62$} & \multicolumn{1}{c||}{$0.063$}  \\[0.2em]
			$800$                      & \multicolumn{1}{c|}{$0.086$} & \multicolumn{1}{c|}{$1.58$} & \multicolumn{1}{c||}{$0.027$}       &\multicolumn{1}{c|}{$0.27$} & \multicolumn{1}{c|}{$7.94$} & \multicolumn{1}{c||}{$0.029$}  \\[0.2em]
			$900$                      & \multicolumn{1}{c|}{$0.042$} & \multicolumn{1}{c|}{$2.98$} & \multicolumn{1}{c||}{$0.013$}       &\multicolumn{1}{c|}{$0.13$} & \multicolumn{1}{c|}{$11.57$} & \multicolumn{1}{c||}{$0.014$}  \\[0.2em]
			$1000$                      & \multicolumn{1}{c|}{$0.021$} & \multicolumn{1}{c|}{$1.84$} & \multicolumn{1}{c||}{$0.0066$}       &\multicolumn{1}{c|}{$0.066$} & \multicolumn{1}{c|}{$11.14$} & \multicolumn{1}{c||}{$0.0069$} \\[0.2em] \hline
			$\green{100}$                      & \multicolumn{1}{c|}{$0.94$} & \multicolumn{1}{c|}{$1.33$} & \multicolumn{1}{c||}{$0.00094$}       &\multicolumn{1}{c|}{$2.98$} & \multicolumn{1}{c|}{$4.22$} & \multicolumn{1}{c||}{$0.0023$} \\[0.2em] 
            $\green{200}$                      & \multicolumn{1}{c|}{$0.25$} & \multicolumn{1}{c|}{$0.36$} & \multicolumn{1}{c||}{$0.00078$}       &\multicolumn{1}{c|}{$0.80$} & \multicolumn{1}{c|}{$1.13$} & \multicolumn{1}{c||}{$0.0023$}  
	\end{tabular}}
	\captionsetup{justification=raggedright,singlelinecheck=false}
	\caption{Signal significance for VLL single-production at the LHC calculated with an evolution algorithm that maximises the Asimov metric. The last two rows represent the singlet model whereas the remaining ten are benchmark points of the doublet model. The colour coding is the same as in \cref{mixings_doublet,mixings_singlet}.}\label{tab:Sig_VLLs}
\end{table}
\begin{table}[htb!]
	\centering
	\resizebox{0.45\textwidth}{!}{\begin{tabular}{c||ccc||}
			\multirow{2}{*}{Mass of VLL (GeV)} & \multicolumn{3}{c||}{$3000$ $\mathrm{fb^{-1}}$}                                                                               \\ \cline{2-4} 
			& $s/\sqrt{s+b}$                          & $\mathcal{Z}(<1\%)$                         & $\mathcal{Z}_A$                
			\\ \hline
			$100$                      & \multicolumn{1}{c|}{$238.17$} & \multicolumn{1}{c|}{$258.55$} & \multicolumn{1}{c||}{$26.13$}   \\[0.2em]
			$200$                      & \multicolumn{1}{c|}{$91.62$} & \multicolumn{1}{c|}{$94.61$} & \multicolumn{1}{c||}{$9.81$}  \\[0.2em]
			$300$                      & \multicolumn{1}{c|}{$25.35$} & \multicolumn{1}{c|}{$25.58$} & \multicolumn{1}{c||}{$2.68$}       \\[0.2em]
			$400$                      & \multicolumn{1}{c|}{$8.73$} & \multicolumn{1}{c|}{$10.19$} & \multicolumn{1}{c||}{$0.92$}
			\\[0.2em]
			$500$                      & \multicolumn{1}{c|}{$3.32$} & \multicolumn{1}{c|}{$11.14$} & \multicolumn{1}{c||}{$0.35$}          \\[0.2em]
			$600$                      & \multicolumn{1}{c|}{$1.37$} & \multicolumn{1}{c|}{$11.45$} & \multicolumn{1}{c||}{$0.14$}         \\[0.2em]
			${\color{red}700}$                      & \multicolumn{1}{c|}{${\color{red}0.61}$} & \multicolumn{1}{c|}{${\color{red}11.62}$} & \multicolumn{1}{c||}{${\color{red}0.063}$}        \\[0.2em]
			$800$                      & \multicolumn{1}{c|}{$0.38$} & \multicolumn{1}{c|}{$11.58$} & \multicolumn{1}{c||}{$0.029$}         \\[0.2em]
			$900$                      & \multicolumn{1}{c|}{$0.21$} & \multicolumn{1}{c|}{$11.54$} & \multicolumn{1}{c||}{$0.014$}        \\[0.2em]
			$1000$                      & \multicolumn{1}{c|}{$0.26$} & \multicolumn{1}{c|}{$11.14$} & \multicolumn{1}{c||}{$0.0069$}        
	\end{tabular}}
	\captionsetup{justification=raggedright,singlelinecheck=false}
	\caption{Signal significance for VLL single-production at the LHC calculated with the neural network trained in the {\color{red}700 GeV} data, with the network details shown in Tab.~\ref{tab:table-VLBSM-EVO-2}}\label{tab:Table_sig_700}
\end{table}
We notice that, both for $\mathcal{L} = 3000$ $\mathrm{fb^{-1}}$ or $\mathcal{L} = 300$ $\mathrm{fb^{-1}}$ as well as any of the mass values, we are able to obtain significances greater than $5\sigma$ if the simplified Asimov metric and $s/\sqrt{s+b}$ are considered. These results may indeed suggest that, significances above 5$\sigma$ can be achievable even for larger VLL mass values at the LHC. This is particularly relevant in the doublet model while for singlet VLLs the significance quickly drops if we go beyond $200~\mathrm{GeV}$.
If only the Asimov significance, $\mathcal{Z}_A$, which is the most conservative one, is considered we can still probe doublet VLLs up to about $200~\mathrm{GeV}$ already at the LHC Run III. For the singlet scenario, considering the HL-LHC program, we can not exclude/discover VLLs, with the hightest significance being 4.22$\sigma$ for the $\mathcal{Z}(<1\%)$ metric. While bellow the discovery threshold, it can still be regarded as a potential anomaly, whose significance can be increased with the combination of additional channels.

Although LEP constraints have already excluded VLLs up to $100.6~\mathrm{GeV}$, our results for $100~\mathrm{GeV}$ are merely indicative of how large can the significance become for small doublet masses, if our analysis technique is employed.

As one can note, there is an overall trend of the significance dropping as the mass of the VLL decreases. However, we note that for masses above 600 GeV, we have consistently obtained signifcances close to $10\sigma$ for the $\mathcal{Z}(<1\%)$ measure. Here, there are two main factors at play. First, considering larger masses, kinematic distributions tend to peak at higher energies, which is particularly relevant for mass distributions, enhancing the neural networks discriminating power. Additionally, we employ the evolutionary algorithm to every single point, meaning that each network is optimized to a specific phase space region\footnote{The neural networks found to be optimized for each point are shown in appendix~\ref{app:Neural_Nets}.}. In a realistic search scenario we would not know the mass of the VLL and therefore it would not be reasonable to set a stronger preference in one of the various networks optimized towards different VLL masses. However, the networks that we use are engineered to be generic enough and can in principle be applied to distinct masses up to a certain discrimination power. To illustrate this, we apply the 700 GeV network to all scanned masses and present the results in \cref{tab:Table_sig_700}. As one can see, the numerical values of the significance changed for the more lenient metrics, $s\sqrt{s+b}$ and $\mathcal{Z}(<1\%)$, whereas for the most conservative one the significance remained the same. However, an increase was also experienced, in particular, for a VLL mass of 800 GeV, where the lenient Asimov metric grew from $7.84\sigma$ to $11.58\sigma$. These results indicate that, despite the networks' training in distinct phase space regions, they are versatile enough to offer a good discrimination power, which also indicates absence of over-fitting. In essence, the hypothetical observation of a statistical excess for a given mass could motivate employing an optimized network to potentially enhance such a signal/excess.

\begin{figure}[!htb]
    \centering
    \captionsetup{justification=raggedright,singlelinecheck=false}
	\includegraphics[width=\textwidth]{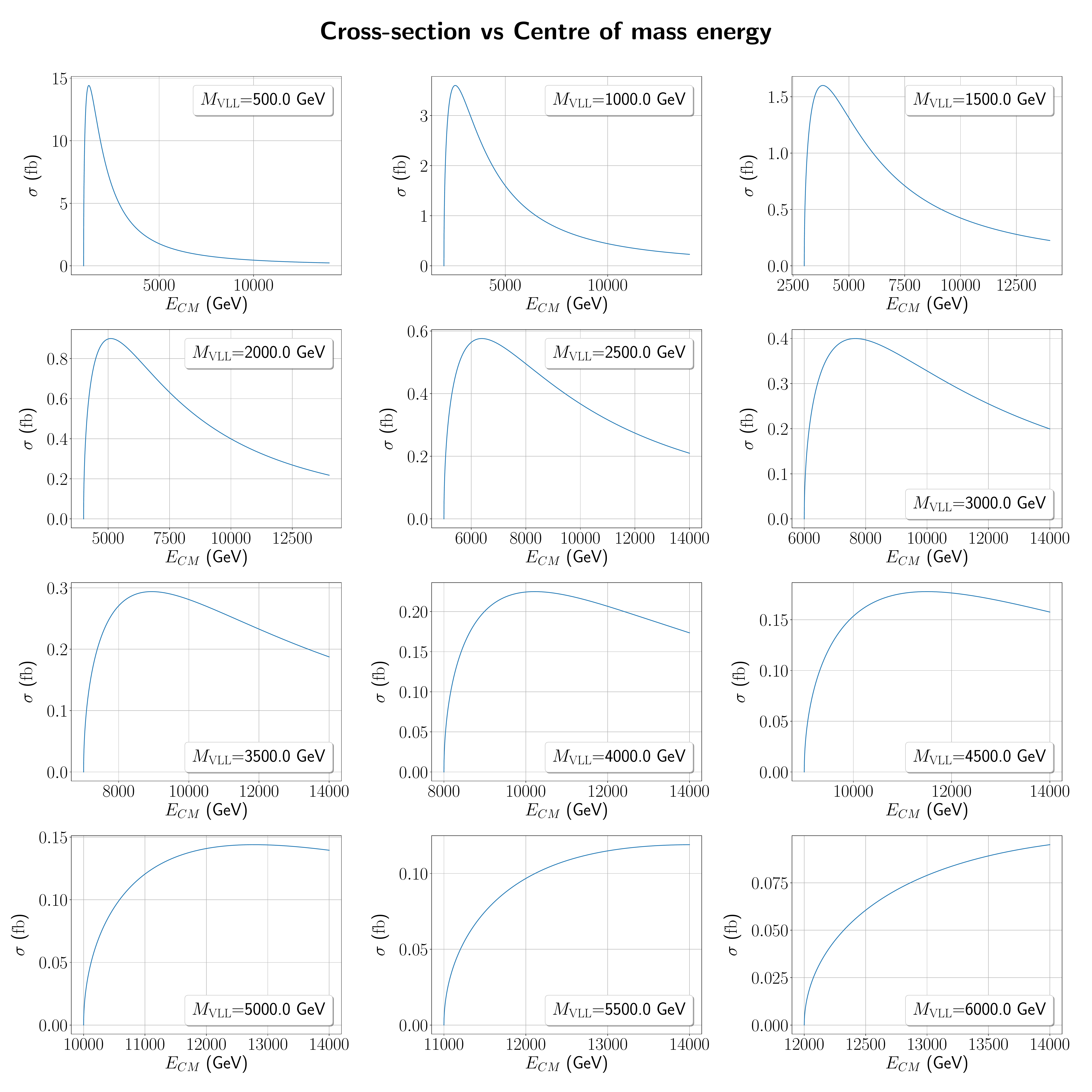}
	\caption{VLL pair production cross section, in fb, as a function of centre-of-mass energy for a lepton collider, in GeV. Each individual box is representative of a different mass of the singlet VLL, which is also given in GeV.}
	\label{fig:Xsec_muon_singlet}
\end{figure}

The low cross-sections of the singlet model call for a different approach on how to probe them. It is in this context that the near future electron and muon colliders can offer new opportunities. As it was mentioned in \cref{subsec:collider_methods}, production of particles via electroweak processes is favoured in these colliders. As such, it is instructive to understand how can a lepton collider enhance the cross-section for the case of pair-produced VLLs via an $s$-channel process. Note that the analysis that follows is independent of the chosen collider, as the $s$-channel cross-section is independent of the mass of the initial colliding particles. Therefore, all results shown here are valid both for the electron and muon machines. At LO, the main diagrams involved are shown in \cref{fig:schannel-muon}. Note that, since we are assuming non-zero couplings between the muon/electron and the VLL, there are also $t$-channel contributions of the form
\begin{equation}\label{eq:TChannel_VLL} \raisebox{-5.2em}{\includegraphics{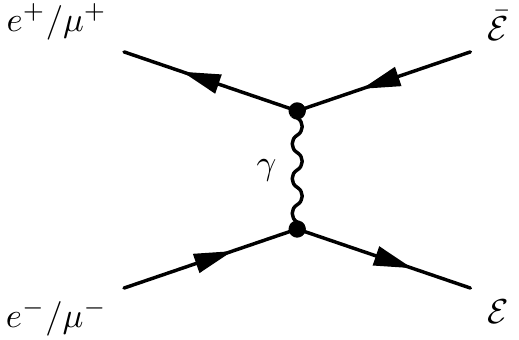}} \quad+\quad
\raisebox{-5.2em}{\includegraphics{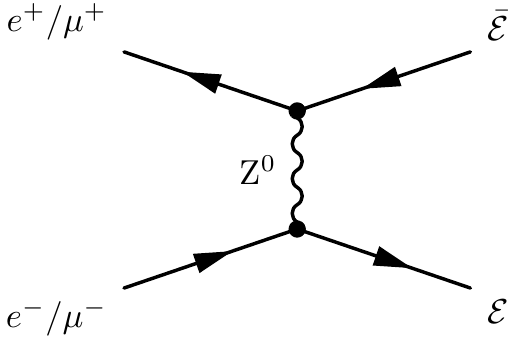}}.
\end{equation}
However, such contributions are sub-leading when compared to production via the $s$-channel process, since these depend on mixing structures of the VLL with the muon, whereas in the $s$-channel the interaction vertices feature two leptons of the same flavour coupling directly to vectors via gauge interactions. Furthermore, in the particular case of our models, the approximate family symmetry that only allows couplings between muons and VLLs results in vanishing $t$-channel contributions at electron colliders, but not at muon machines. Note that this channel can be seen as a direct probe to the flavour structure in the leptonic sector if both muon and electron colliders become operational. It is then safe to neglect the $t$-channel contribution in the remainder of this analysis, where we use the same mixing structure as defined in \cref{mixings_singlet}. We then consider the $ \mathcal{E} \to W\nu_\ell$ channel, which, depending on how the $W$ bosons decay, can lead to the $2\nu_\ell + 4j$ or $2\nu_\ell + 2j + \ell\nu_\ell$ or $2\nu_\ell + \ell\nu_\ell + \ell\nu_\ell$ final states. For the topologies that involve jets as final states, the main backgrounds include diboson production ($WW$, $W\mathrm{Z^0}$ and $\mathrm{Z^0Z^0}$) with associated jets as well as tau pair production. For the purely leptonic final state $2\nu_\ell + \ell\nu_\ell + \ell\nu_\ell$, diboson production is relevant. Indeed, a lepton collider is a much cleaner environment when compared to a hadronic machines in such a way that backgrounds involving jets can be safely discarded. VLLs can also decay into $\mathrm{Z^0}$ bosons as $E(\mathcal{E})\rightarrow \mu \mathrm{Z^0}$. Such a channel would be an important test for a hypothetical VLL discovery provided that the final states can contain at least 6 charged leptons. Besides being a very clean process, 6 lepton topologies are also expected to be small in the context of a SM background that typically results from triboson production processes. A detailed phenomenological analysis at lepton colliders using analogous deep learning methods to those discussed in the context of the LHC is beyond the scope of this article and is left for future work.

For a comprehensive understanding on how the cross section depends on a lepton collider center-of-mass energy, $E_\mathrm{CM}$, we use \texttt{FeynCalc} to obtain the following expression
\begin{equation}\label{eq:xsec_Analytical}
\begin{aligned}
\sigma = \dfrac{\sqrt{\dfrac{E_{\mathrm{CM}}^2}{4}- M_\mathcal{E}^2}}{256\pi E_{\mathrm{CM}}^7(M_{\mathrm{Z^0}}^2 - E_{\mathrm{CM}}^2)^2}(&\alpha_1 E_{\mathrm{CM}}^8 + \alpha_2 E_{\mathrm{CM}}^6 M_{\mathcal{E}}^2 +\alpha_3E_{\mathrm{CM}}^6 M_{\mathrm{Z^0}}^2 + \alpha_4E_{\mathrm{CM}}^4 M_{\mathrm{Z^0}}^2M_{\mathcal{E}}^2 + \alpha_5 E_{\mathrm{CM}}^4 M_{\mathrm{Z^0}}^4 +\\ & +\alpha_6E_{\mathrm{CM}}^2 M_{\mathrm{Z^0}}^4M_{\mathcal{E}}^2),
\end{aligned}
\end{equation}
where $\alpha_i$ for $i=1,\dots,6$ are dimensionless constants, proportional to the product of various couplings, i.e.~$\alpha_j = \alpha_j(U_\mathrm{L}^e, U_\mathrm{R}^e, g, g', \theta_W)$, $M_{\mathrm{Z^0}}$ is the mass of the $\mathrm{Z^0}$ boson and $M_{\mathcal{E}}$ the mass of the singlet VLL. With the numerical values in \cref{mixings_singlet} these constants read as
\begin{equation}\label{eq:alpha_constants}
\begin{aligned}
\alpha_1 = 0.189325, \quad \alpha_2 = -0.213435, \quad \alpha_3 = -0.328608, \quad \alpha_4 = 0.403093, \quad \alpha_5 = 0.152126, \quad \alpha_6 = -0.209655.
\end{aligned}
\end{equation}
In \cref{fig:Xsec_muon_singlet} we plot the corresponding cross-section as a function of the centre-of-mass energy, for VLL masses in the range between $500~\mathrm{GeV}$ and $6~\mathrm{TeV}$. Do note that we are studying the singlet scenario, where the decay width always remain bellow the mass of the VLL. It is interesting to note that the cross sections are rather large, above 13.23~$\mathrm{fb}$ for a VLL mass of $500~\mathrm{GeV}$ and $E_{\mathrm{CM}} = 1.5$ TeV, dropping to $0.095~\mathrm{fb}$ for a VLL mass of $6\mathrm{TeV}$ and $E_{\mathrm{CM}} = 14$~TeV. This increase of the cross section allows to probe higher mass ranges than the ones at the reach of the LHC, in a much cleaner environment. Immediately noticeable is the fact that the cross section hits a maximum shortly after $E_{\mathrm{CM}} \sim 2 M_{\mathrm{VLL}}$, with a subsequent drop as $E_{\mathrm{CM}}$ increases. This implies that when the collider beam energy is two times the mass of the VLL, the double production of VLL is enhanced and the discovery potential is maximized. For lower masses, the drop is more pronounced when compared to the high-mass regime, essentially because we are taking $E_\mathrm{CM}$ in the range of $3 - 14$~TeV for low masses, while for high masses it occurs for much higher centre-of-mass energies (beyond $E_{\mathrm{CM}} = 14$ TeV), and therefore not as relevant for the proposed lepton colliders. In particular, for $M_{\rm VLL}=500$ GeV and at $E_{\mathrm{CM}} = 14$ TeV we have $\sigma = 0.23$ $\mathrm{fb}$, while for the same mass, at $E_{\mathrm{CM}} = 3$~TeV, we have $\sigma = 4.63$ $\mathrm{fb}$. 
\begin{table}[]
\resizebox{\textwidth}{!}{\begin{tabular}{c||c|c|c|c|c|c|c|c||}
                    & \green{200} GeV                 & \green{700} GeV                 & \green{1200} GeV                & \green{1700} GeV                & \green{2200} GeV                & \green{2700} GeV                & \green{3200} GeV                & \green{3700} GeV                                     \\ \hline
$E_{\mathrm{CM}} = 1.5$ TeV  & 19.18 $\mathrm{fb}$ & 5.51 $\mathrm{fb}$ & $-$ & $-$                     & $-$                     & $-$                     & $-$                     & \multicolumn{1}{c||}{$-$}                     \\
$E_{\mathrm{CM}} = 3$ TeV  & 5.00 $\mathrm{fb}$ & 4.21 $\mathrm{fb}$ & 2.49 $\mathrm{fb}$ & $-$                     & $-$                     & $-$                     & $-$                     & \multicolumn{1}{c||}{$-$}                     \\
$E_{\mathrm{CM}} = 10$ TeV & 0.46 $\mathrm{fb}$ & 0.45 $\mathrm{fb}$ & 0.44 $\mathrm{fb}$ & 0.41 $\mathrm{fb}$ & 0.39 $\mathrm{fb}$ & 0.35 $\mathrm{fb}$ & 0.31 $\mathrm{fb}$ & \multicolumn{1}{c||}{0.26 $\mathrm{fb}$} \\
$E_{\mathrm{CM}} = 14$ TeV & 0.23 $\mathrm{fb}$ & 0.23 $\mathrm{fb}$ & 0.23 $\mathrm{fb}$ & 0.22 $\mathrm{fb}$ & 0.21 $\mathrm{fb}$ & 0.21 $\mathrm{fb}$ & 0.19 $\mathrm{fb}$ & \multicolumn{1}{c||}{0.18 $\mathrm{fb}$} \\
\hline
$E_{\mathrm{CM}} = 14$ TeV ($e^+ e^-\gamma\gamma$) & 17.95 $\mathrm{fb}$ & 0.54 $\mathrm{fb}$ & 0.095 $\mathrm{fb}$ & 0.027 $\mathrm{fb}$ & 0.0095 $\mathrm{fb}$ & 0.0037 $\mathrm{fb}$ & 0.001585 $\mathrm{fb}$ & \multicolumn{1}{c||}{0.00069 $\mathrm{fb}$}\\
$E_{\mathrm{CM}} = 14$ TeV ($\mu^+ \mu^-\gamma\gamma$) & 9.10 $\mathrm{fb}$ & 0.26 $\mathrm{fb}$ & 0.045 $\mathrm{fb}$ & 0.012 $\mathrm{fb}$ & 0.0043 $\mathrm{fb}$ & 0.0017 $\mathrm{fb}$ & 0.00070 $\mathrm{fb}$ & \multicolumn{1}{c||}{0.00030 $\mathrm{fb}$}
\end{tabular}}
\captionsetup{justification=raggedright,singlelinecheck=false}
\caption{Singlet VLL double production cross-section at lepton colliders. As idicated in the top row of the table, all considered mass values correspond to the green benchmark point in \cref{mixings_singlet}. We show the cross-section for a centre-of-mass energy $E_{\mathrm{CM}} = 1.5$ and 3 TeV, which are the target energies for the CLIC collider, whereas $E_{\mathrm{CM}} = 3$, 10 and 14 TeV correspond to the current proposal for the future $\mu^+\mu^-$ collider. Points marked with~``$-$'' indicate that there is not enough energy to pair-produce the particles at that mass.}
\label{tab:Muon-CLIC-numbers}
\end{table}
Fixing the centre-of-mass energy and looking at various mass points, as shown in Tab.~\ref{tab:Muon-CLIC-numbers}, we notice that for the proposed high-energy colliders, the variation of the mass does not cause significant deviations in the cross-section. For example, taking $E_{\mathrm{CM}} = 14$ TeV, we observe that the cross-section remains nearly constant for the displayed masses, ranging from $0.23~\mathrm{fb}$ for a 200 GeV VLL to $0.18~\mathrm{fb}$ for a $3.7~\mathrm{TeV}$ one. 

We end this section with a comment about the cross-sections at the high energy end of the colliders. As the energy grows the $s$-channel cross sections decrease. However, there is an alternative process that grows with $\ln^2 (s/m_f^2)$, where $f$ stands for the incoming fermion.
This is $e^+ e^- \to e^+ e^-  E({\mathcal{E}}) \bar E (\bar {\mathcal{E}})$ for an electron-positron collider and $\mu^+ \mu^- \to \mu^+ \mu^-  E({\mathcal{E}}) \bar E (\bar {\mathcal{E}}) $ for a muon collider. The photon fusion processes have cross-sections that grow
with the centre-of-mass energy~\cite{Budnev:1975poe, Frixione:1993yw} and although they are not competitive for the lower energies they become dominant at high energies. We present in Tab.~\ref{tab:Muon-CLIC-numbers} the values of these cross-sections for an energy of 14 TeV
which shows that they can play an important role for very high energy lepton colliders.

\section{Conclusions}\label{sec:Conclusions}

In this article we have studied two simple SM extensions featuring, each of them, a new vector-like lepton and a sterile neutrino. We have confronted the cases of a doublet and a singlet VLL and discussed their collider phenomenology, both at the LHC and future leptonic machines. For the former we have employed Deep Learning techniques to compute the statistical significance of a hypothetical discovery. In the selection of benchmark scenarios we have required that the coupling of VLL to muons is consistent with flavour constraints and the branching fractions of Higgs and Z bosons to muons. We have also shown that, decay width of both VLL doublet components becomes increasingly larger with growing exotic lepton masses. In our analysis we have only considered scenarios where the width is smaller than the mass.

In the context of the LHC studies, we have performed Monte-Carlo simulations to generate data for signal and background topologies. The signal is characterized by the presence of a single isolated lepton and a substantial amount of MET for the singlet case. For the doublet case, the signal involves 4 light jets, a charged lepton and a neutrino as final state particles. To separate the signal from the background we constructed neural networks which follow from an implementation of an evolution algorithm that maximises the Asimov significance. We have shown that, for doublet VLLs, we can  exclude masses up to 1 TeV with more than five standard deviations. In particular, for masses of $1~\mathrm{TeV}$, we obtain a significance of $\mathcal{Z}(<1\%) = 11.14\sigma$ for the high luminosity phase of the LHC, with an integrated luminosity of $\mathcal{L} = 3000$ $\mathrm{fb^{-1}}$. We have also verified that one can already test the doublet VLL scenario at the Run III of the LHC, which will deliver $\mathcal{L} = 300$ $\mathrm{fb^{-1}}$ of data. For such luminosity, one can test VLL masses up to about $300~\mathrm{GeV}$ obtaining $s/\sqrt{s+b} = 8.02\sigma$, $\mathcal{Z}(<1\%) = 8.09\sigma$ and $\mathcal{Z}_A = 2.56\sigma$. For the singlet scenario, production cross-sections are substantially smaller and the expected number of events is usually zero, with an exception for VLL masses of $100~\mathrm{GeV}$ and $200~\mathrm{GeV}$. While for the former one finds $s/\sqrt{s+b} = 2.98\sigma$, $\mathcal{Z}(<1\%) = 4.22\sigma$ and $\mathcal{Z}_A = 0.0023\sigma$ at $\mathcal{L} = 3000$ $\mathrm{fb^{-1}}$, the run III estimation gives $s/\sqrt{s+b} = 0.94\sigma$, $\mathcal{Z}(<1\%) = 1.33\sigma$ and $\mathcal{Z}_A = 0.00094\sigma$ at $\mathcal{L} = 300$ $\mathrm{fb^{-1}}$, experiencing a great drop for a mass of $200~\mathrm{GeV}$. In particular, the latter scenario can not be excluded, with a statistical significance bellow the discovery threshold.

Owning to the low production cross-sections of the singlet scenario, a supplementary analysis was made within the context of lepton colliders. We have performed numerical computations for the expected VLL pair-production cross-section in the $s$-channel. We find that, in general, larger cross-sections are obtained when compared to the LHC analysis, allowing for a much wider range of masses to be probed with relevance for singlet VLLs. In particular, we note that for luminosities of the order of $\mathrm{ab}^{-1}$ the, even for singlet VLLs with mass $3.7~\mathrm{TeV}$ one can expect $260$ events at $E_{\mathrm{CM}} = 10$ TeV and $180$ events if $E_{\mathrm{CM}} = 14$ TeV. Furthermore, even for a VLL as heavy as $6~\mathrm{TeV}$, a $14~\mathrm{TeV}$ lepton machine delivering a luminosity of $1~\mathrm{ab}^{-1}$ is expected to produce almost 100 events. With this in mind we conclude that the study of VLL particles at future electron or muon colliders is a rather relevant physics case scenario to be explored, allowing to significantly extend the current reach of the LHC.  \\ \\
{\bf Acknowledgements.}
The authors would like to thank Celso Nishi for valuable comments made to the initial version of this manuscript. J.G., F.F.F., and A.P.M. are supported by the Center for Research and Development in Mathematics and Applications (CIDMA) through the Portuguese Foundation for Science and Technology (FCT - Funda\c{c}\~{a}o para a Ci\^{e}ncia e a Tecnologia), references UIDB/04106/2020 and UIDP/04106/2020. A.P.M., F.F.F., J.G. and R.S. are supported by the project PTDC/FIS-PAR/31000/2017. A.P.M., F.F.F., J.G. are also supported by the projects CERN/FIS-PAR/0021/2021. Additionally, A.P.M. and  J.G. are supported by CERN/FIS-PAR/0019/2021.
J.G. is also directly funded by FCT through the doctoral program grant with the reference 2021.04527.BD.
A.P.M.~is also supported by national funds (OE), through FCT, I.P., in the scope of the framework contract foreseen in the numbers 4, 5 and 6 of the article 23, of the Decree-Law 57/2016, of August 29, changed by Law 57/2017, of July 19.
R.P.~is supported in part by the Swedish Research Council grant, contract number 2016-05996, as well as by the European Research Council (ERC) under the European Union's Horizon 2020 research and innovation programme (grant agreement No 668679).
R.S. is supported by CFTC-UL under FCT contracts UIDB/00618/2020, UIDP/00618/2020, and by the projects CERN/FISPAR /0002/2017, CERN/FIS-PAR/0014/2019 and by the HARMONIA project of the National Science Centre, Poland, under contract UMO- 2015/18/M/ST2/00518. A.O. is supported by the FCT project CERN/FIS-PAR/0029/2019.

\appendix

\section{Feynman Rules}\label{app:Feyn_rules}

In this appendix, the list of Feynman rules relevant for the numerical analysis is presented. All rules are displayed in the mass basis. To simplify the notation, we define the chirality projection operators as $\mathrm{P_L} = (1-\gamma_5)/2$ and $\mathrm{P_R} = (1+\gamma_5)/2$. Notation wise, we define the following: $\theta_W$ is the Weinberg mixing angle, $g$ the $\SU{2}{L}$ gauge coupling and $g'$ the $\U{1}{Y}$ gauge coupling. The Latin indices $i,j = (1,2,3,4)$ denote $(e, \mu, \tau, E(\mathcal{E}))$ for leptons, while for neutrinos $i,j = (1,2,3,4,5,6) = (\nu_1, \nu_2 ,\nu_3, \nu_4, \nu_{5}, \nu_{6})$, with $\nu_5$ and $\nu_6$ arising from the doublet representation of the VLL and $\nu_4$ is the sterile neutrino. For the singlet model, there is no extra left-handed neutrinos, hence the latin indices only run from 1 to 4.

For the doublet case, the Feynman rules read as
\\ \\
\begin{equation}\label{eq:Feynman_Rules_Zboson}
\begin{aligned}
&\hspace{2.01cm}\raisebox{-3.35em}{\includegraphics{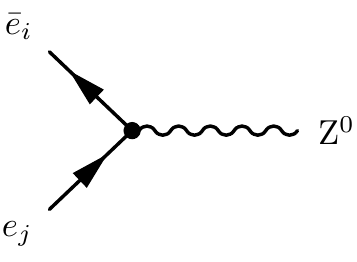}} = i\qty(g^Z_L \gamma_\mu\mathrm{P_L} + g^Z_R \gamma_\mu\mathrm{P_R}), \\
&g_L^Z = \frac{1}{2}\qty(g\cos{\theta_W} - g'\sin{\theta_W})\qty[\qty(U_L^e)^*_{j4}\qty(U_L^e)_{i4} + \sum_{a=1}^3\qty(U_L^e)^*_{ja}\qty(U_L^e)_{ia}], \\
&g_R^Z = -g'\sin(\theta_W)\sum_{a=1}^3\qty(U_R^e)^*_{ia}\qty(U_R^e)_{ja} + \frac{1}{2}\qty(U_R^e)^*_{i4}\qty(U_R^e)_{j4}\qty(g\cos{\theta_W} - g'\sin{\theta_W}),
\end{aligned}
\end{equation}

\begin{equation}\label{eq:Feynman_Rules_Wplus}
\begin{aligned}
&\hspace{0.25cm}\raisebox{-3.35em}{\includegraphics{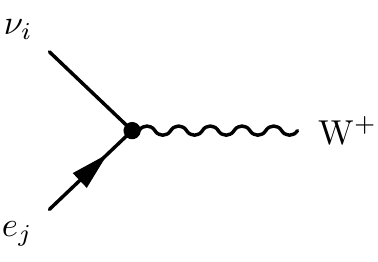}} = i\qty(g^{W^+}_L \gamma_\mu\mathrm{P_L} + g^{W^+}_R\gamma_\mu \mathrm{P_R}), \\
&g_L^{W^+} = -\frac{g}{\sqrt{2}} \qty[\qty(U_L^e)^*_{j4}\qty(U_\nu)_{i5} + \sum_{a=1}^3\qty(U_L^e)^*_{ja}\qty(U_\nu)_{ia}], \\ 
&g_R^{W^+} = -\frac{g}{\sqrt{2}}\qty(U_\nu)^*_{i6}\qty(U_R^e)_{j4},
\end{aligned}
\end{equation}

\begin{equation}\label{eq:Feynman_Rules_Wminus}
\begin{aligned}
&\hspace{0.25cm}\raisebox{-3.35em}{\includegraphics{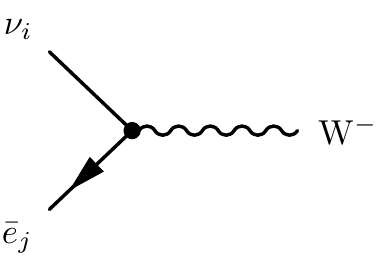}} = i\qty(g^{W^-}_L \gamma_\mu\mathrm{P_L} + g^{W^-}_R \gamma_\mu\mathrm{P_R}), \\
&g_L^{W^-} = -\frac{g}{\sqrt{2}} \qty[\qty(U_\nu)^*_{i5}\qty(U_L^e)_{j4} + \sum_{a=1}^3\qty(U_\nu)^*_{ia}\qty(U_L^e)_{ja}], \\
&g_R^{W^-} = -\frac{g}{\sqrt{2}}\qty(U_R^e)^*_{j4}\qty(U_\nu)_{i6},
\end{aligned}
\end{equation}

\begin{equation}\label{eq:Feynman_Rules_Higgs}
\begin{aligned}
&\hspace{0.85cm}\raisebox{-3.35em}{\includegraphics{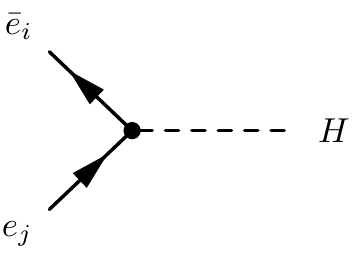}} = i\qty(g^{H}_L \mathrm{P_L} + g^{H}_R \mathrm{P_R}), \\
&g_L^{H} = \frac{1}{\sqrt{2}} \qty[\sum_{a=1}^3\qty(U_L^e)^*_{j4}\qty(U_R^e)^*_{ia}\Theta_a + \sum_{a=1}^3\sum_{b=1}^3\qty(U_R^e)^*_{ib}\qty(U_L^e)^*_{ja}\Pi_{ab}], \\
&g_R^{H} = \frac{1}{\sqrt{2}} \qty[\sum_{a=1}^3\qty(U_R^e)_{ja}\qty(U_L^e)_{i4}\Theta_a^* + \sum_{a=1}^3\sum_{b=1}^3\qty(U_L^e)_{ia}\qty(U_R^e)_{jb}\Pi_{ab}^*],
\end{aligned}
\end{equation}

Similarly, for the singlet case we have,
\\ \\
\begin{equation}\label{eq:Feynman_Rules_Zboson_1}
\begin{aligned}
&\hspace{1.95cm}\raisebox{-3.35em}{\includegraphics{ZBoson-doublet.pdf}} = i\qty(g^Z_L\gamma_\mu \mathrm{P_L} + g^Z_R\gamma_\mu \mathrm{P_R}), \\
&g_L^Z = -g' \qty(U^e_L)^*_{j4}\qty(U^e_L)_{i4}\sin(\theta_W) + \frac{1}{2}\qty(g\cos{\theta_W} - g'\sin{\theta_W}) \sum_{a=1}^3\qty(U_L^e)^*_{ja}\qty(U_L^e)_{ia}, \\
&g_R^Z = -g'\sin(\theta_W)\qty[\qty(U_R^e)^*_{i4}\qty(U_R^e)_{j4} + \sum_{a=1}^3\qty(U_R^e)^*_{ia}\qty(U_R^e)_{ja}],
\end{aligned}
\end{equation}

\begin{equation}\label{eq:Feynman_Rules_Wplus_1}
\begin{aligned}
&\hspace{-1.38cm}\raisebox{-3.35em}{\includegraphics{Wplus-doublet.pdf}} = i g^{W^+}_L\gamma_\mu \mathrm{P_L}, \\
&g_L^{W^+} = -\frac{g}{\sqrt{2}} \sum_{a=1}^3\qty(U_L^e)^*_{ja}\qty(U_\nu)_{ia},
\end{aligned}
\end{equation}

\begin{equation}\label{eq:Feynman_Rules_Wminus_2}
\begin{aligned}
&\hspace{-1.38cm}\raisebox{-3.35em}{\includegraphics{Wminus-doublet.pdf}} = ig^{W^-}_L\gamma_\mu \mathrm{P_L}, \\
&g_L^{W^-} = -\frac{g}{\sqrt{2}} \sum_{a=1}^3\qty(U_\nu)^*_{ia}\qty(U_L^e)_{ja}, 
\end{aligned}
\end{equation}

\begin{equation}\label{eq:Feynman_Rules_Higgs_1}
\begin{aligned}
&\hspace{0.80cm}\raisebox{-3.35em}{\includegraphics{Higgs-doublet.pdf}} = i\qty(g^{H}_L \mathrm{P_L} + g^{H}_R \mathrm{P_R}), \\
&g_L^{H} = \frac{1}{\sqrt{2}} \qty[\sum_{a=1}^3\qty(U_R^e)^*_{i4}\qty(U_L^e)^*_{ja}\theta_a + \sum_{a=1}^3\sum_{b=1}^3\qty(U_R^e)^*_{ib}\qty(U_L^e)^*_{ja}\Pi_{ab}], \\
&g_R^{H} = \frac{1}{\sqrt{2}} \qty[\sum_{a=1}^3\qty(U_L^e)_{ia}\qty(U_R^e)_{j4}\theta_a^* + \sum_{a=1}^3\sum_{b=1}^3\qty(U_L^e)_{ia}\qty(U_R^e)_{jb}\Pi_{ab}^*].
\end{aligned}
\end{equation}

\cleardoublepage

\section{Kinematic distributions for VLL single-production}\label{app:Kinematic_dist}

\begin{figure}[htb!]
    \centering
    \captionsetup{justification=raggedright,singlelinecheck=false}
	\includegraphics[width=\textwidth,height=\textheight,keepaspectratio]{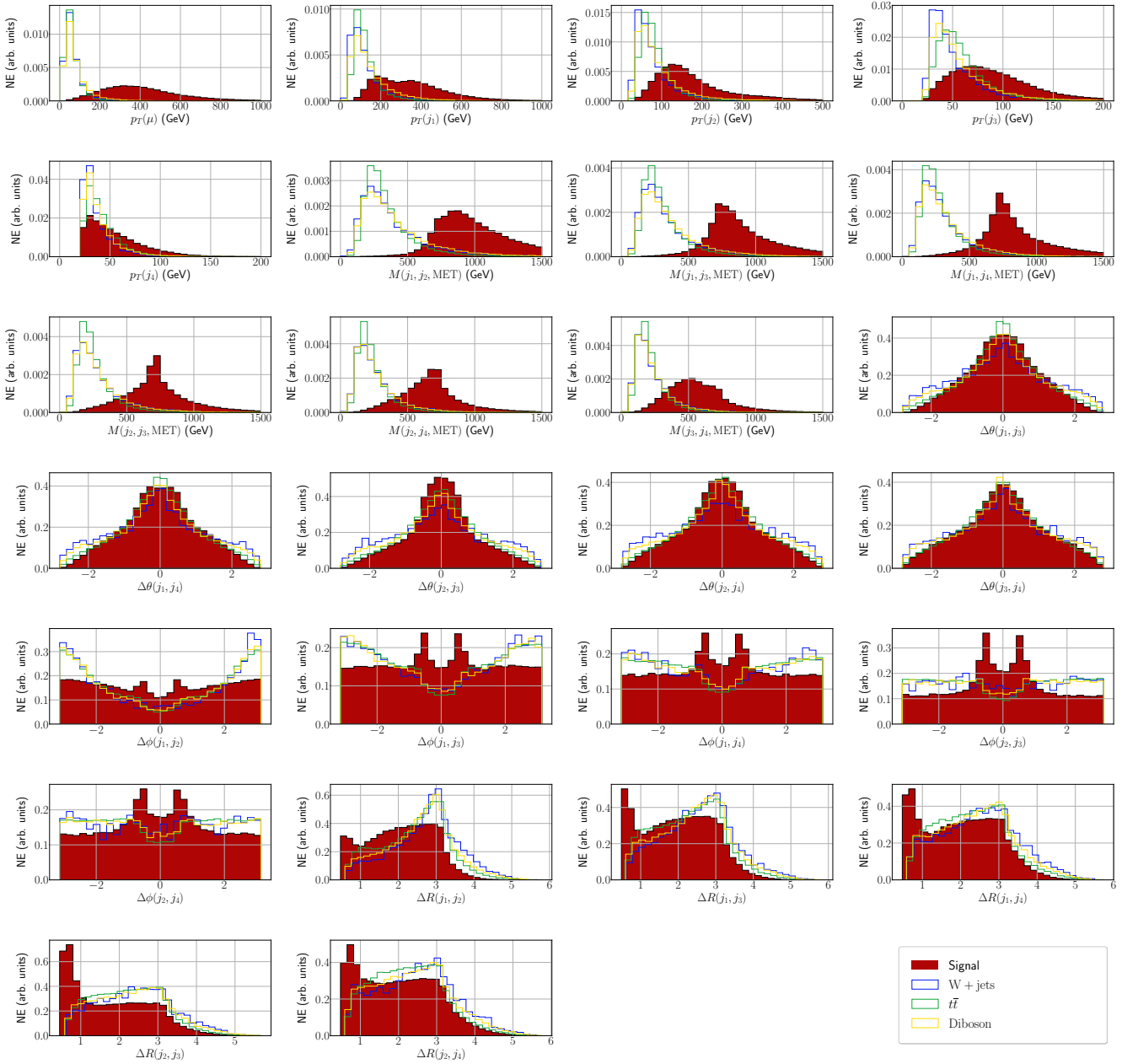}
	\caption{A sample of the kinematic variables that the neural network uses for classification. Data is simulated for a doublet VLL with mass of 700 GeV and in the ATLAS detector. From left to right and top to bottom, the variables are $p_T$ of the muon, $p_T$ of jet 1, 2, 3 and 4, mass distributions for the combinations $(j_1, j_2, \mathrm{MET})$, $(j_1, j_3, \mathrm{MET})$, $(j_1, j_4, \mathrm{MET})$, $(j_2, j_3, \mathrm{MET})$, $(j_2, j_4, \mathrm{MET})$, $(j_3, j_4, \mathrm{MET})$, $\Delta \theta$ distributions for the combinations $(j_1, j_3)$, $(j_1, j_4)$, $(j_2, j_3)$, $(j_2, j_4)$ and $(j_3, j_4)$, $\Delta \phi$ distributions for the combinations $(j_1, j_2)$, $(j_1, j_3)$, $(j_1, j_4)$, $(j_2, j_3)$, $(j_2, j_4)$ and $\Delta R$ distributions for the combinations $(j_1,j_2)$, $(j_1,j_3)$, $(j_1,j_4)$, $(j_2,j_3)$ and $(j_2,j_4)$. In the $y$-axis, it is indicated that events are normalized (NE). We consider 30 bins for all background and signal histograms.}
	\label{fig:Doublet_kin}
\end{figure}

\begin{figure}[htb!]
    \centering
    \captionsetup{justification=raggedright,singlelinecheck=false}
	\includegraphics[width=\textwidth,height=\textheight,keepaspectratio]{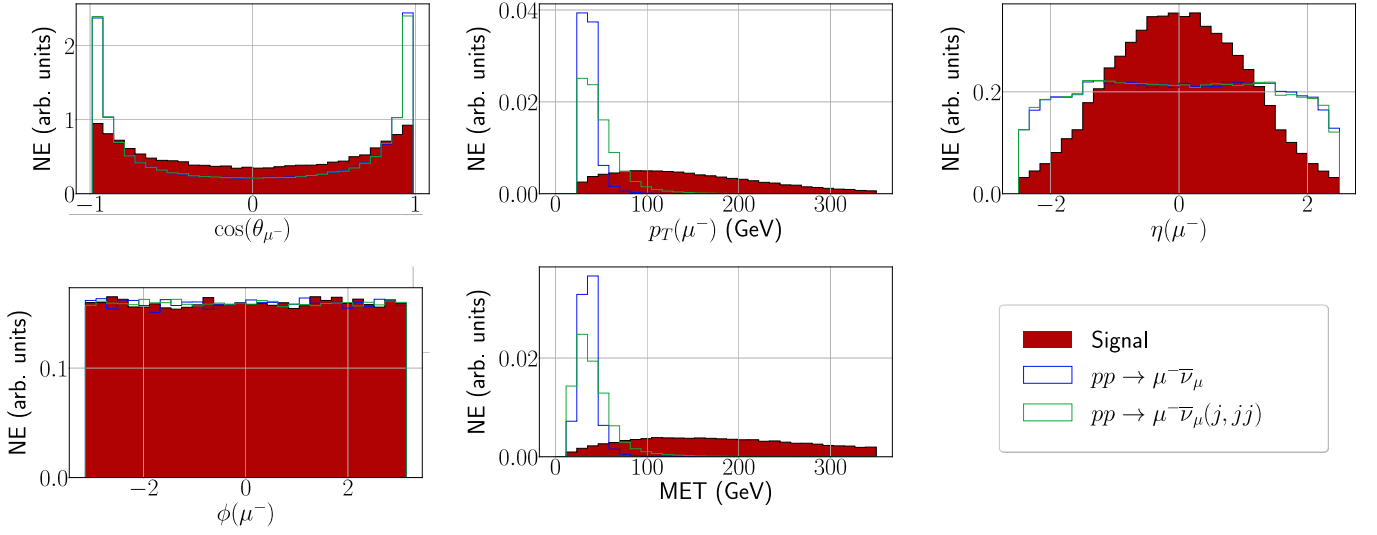}
	\caption{Kinematic variables that the neural network uses for classification. Data is simulated for a doublet VLL with mass of 700 GeV and in the ATLAS detector. From left to right and top to bottom, the variables are $\cos(\theta_{\mu^-})$, transverse momentum of the muon, pseudo-rapidity of the muon, azimuthal angle of the muon and MET. In the $y$-axis, it is indicated that events are normalized (NE). We consider 30 bins for all background and signal histograms.}
	\label{fig:Singlet_kin}
\end{figure}

\clearpage

\section{Neural networks found via the genetic algorithm}\label{app:Neural_Nets}

\begin{table*}[ht!]    
	\centering
    \captionsetup{justification=raggedright,singlelinecheck=false}
	\resizebox{0.91\textwidth}{!}{\begin{tabular}{|c|c|}
		\hline
    	Mass	& Doublet \\
		\hline
		\hline
		\midrule
		\makecell{100 GeV}  & \makecell{\underline{Layers} : 1 input + 1 hidden +1 output. \\ Input and hidden layer with 1024 neurons, \\ output layer with 4 neurons \\
		\underline{Regularizer} : L1L2 (layer 1 and 2) and none (layer 3)\\
		\underline{Activation function} : RELU (for layers 1 and 2) and Sigmoid (layer 3) \\
		\underline{Initializer} : VarianceScaling, with uniform distribution \\
		(layer 1 and 2) in fan\_in mode \\ and GlorotUniform (layer 3) with null seed \\
		}  \\
		\hline
		\makecell{200 GeV} & \makecell{\underline{Layers} : 1 input + 3 hidden + 1 output. \\Hidden and input layers with 1024 neurons each, \\ output layer with 4 neurons \\
		\underline{Regularizer} : L1L2 (layers 1 to 4) and none (for layer 5)\\
		\underline{Activation function} : RELU (layers 1 to 4) and Sigmoid (layer 5)  \\
		\underline{Initializer} : VarianceScaling, with uniform distribution \\
		(layers 1-4) in fan\_in mode \\ and GlorotUniform (layer 5) with null seed \\
		} \\
		\hline
		\makecell{300 GeV}  & \makecell{\underline{Layers} : 1 input + 4 hidden + 1 output. \\ Input and hidden layers with 512 neurons, \\ output layer with 4 neurons \\
		\underline{Regularizer} : L1L2 (for layers 1-5) and none (for layer 6)\\
		\underline{Activation function} : Sigmoid (for all layers) \\
		\underline{Initializer} : VarianceScaling, with truncated normal distribution \\
		(for layers 1-5) in fan\_in mode and \\ GlorotUniform distribution (layer 6) with null seed \\
		} \\
		\hline
		\makecell{400 GeV}  & \makecell{\underline{Layers} : 1 input + 2 hidden + 1 output. \\Hidden and input layers with 256 neurons each, \\ output layer with 4 neurons \\
		\underline{Regularizer} : L1L2 (for layers 1-3) and none (for layer 4)\\
		\underline{Activation function} : RELU (layers 1 to 3) and Sigmoid (layer 4)\\
		\underline{Initializer} : RandomNormal, with 0 mean and 0.05 standard deviation \\ (for layers 1 to 3) and GlorotUniform distribution (layer 4) with null seed\\
		} \\
		\hline
		\makecell{500 GeV}  & \makecell{\underline{Layers} : 1 input + 1 output. \\Input layer with 1024 neurons, \\ output layer with 4 neurons \\
		\underline{Regularizer} : L1L2 (layer 1) and  none (layer 2)\\
		\underline{Activation function} : Sigmoid (all layers)  \\
		\underline{Initializer} : VarianceScaling, with truncated normal distribution 
		(layer 1) \\ in fan\_in mode and GlorotUniform distribution (layer 2) with null seed \\
		} \\
		\hline
	\end{tabular}}
	\caption{Architectures for the neural networks that the genetic algorithm finds for the scanned masses (100 to 500 GeV) of the doublet VLL scenario.}
	\label{tab:table-VLBSM-EVO-1}
\end{table*}

\begin{table*}[ht!]    
	\centering
    \captionsetup{justification=raggedright,singlelinecheck=false}
	\resizebox{0.91\textwidth}{!}{\begin{tabular}{|c|c|}
		\hline
    	Mass	& Doublet \\
		\hline
		\hline
		\midrule
		\makecell{600 GeV}  & \makecell{\underline{Layers} : 1 input + 4 hidden + 1 output. \\Hidden and input layers with 512 neurons each, \\ output layer with 4 neurons \\
		\underline{Regularizer} : L1L2 (for layers 1 to 5) and none (for layer 6)\\
		\underline{Activation function} : Sigmoid (for all layers) \\
		\underline{Initializer} : RandomNormal, with mean 0 and 0.05 standard deviation \\
		(for layers 1 to 5) and GlorotUniform distribution (layer 6) with null seed \\
		}  \\
		\hline
		\makecell{700 GeV} & \makecell{\underline{Layers} : 1 input + 3 hidden + 1 output. \\Hidden and input layers with 512 neurons each, \\ output layer with 4 neurons \\
		\underline{Regularizer} : L1L2 (for layers 1 to 4) and none (for layer 5)\\
		\underline{Activation function} : Sigmoid (for all layers)  \\
		\underline{Initializer} : VarianceScaling, with uniform distribution \\ (for layers 1 to 4) in fan\_in mode and \\ GlorotUniform distribution (layer 5) with null seed\\
		} \\
		\hline
		\makecell{800 GeV}  & \makecell{\underline{Layers} : 1 input + 3 hidden + 1 output. \\Hidden and input layers with 256 neurons each, \\ output layer with 4 neurons \\
		\underline{Regularizer} : L1L2 (for layers 1 and 4) and none (for layer 5)\\
		\underline{Activation function} : ELU (for layers 1 to 4) and Sigmoid (layer 5) \\
		\underline{Initializer} : VarianceScaling, with truncated normal distribution \\ (for layers 1 to 4) in fan\_in mode and \\ GlorotUniform distribution (layer 5) with null seed\\
		} \\
		\hline
		\makecell{900 GeV}  & \makecell{\underline{Layers} : 1 input + 1 hidden + 1 output. \\Hidden and input layers with 256 neurons each, \\ output layer with 4 neurons \\
		\underline{Regularizer} : L1L2 (for layers 1 and 2) and none (for layer 3)\\
		\underline{Activation function} : Tanh (for layers 1 and 2) and Sigmoid (layer 3) \\
		\underline{Initializer} : RandomNormal, with mean 0 and 0.05 standard deviation \\
		(for layers 1 and 2) and GlorotUniform distribution (layer 3) with null seed\\
		} \\
		\hline
		\makecell{1000 GeV}  & \makecell{\underline{Layers} : 1 input + 2 hidden + 1 output. \\Hidden and input layers with 2048 neurons each, \\ output layer with 4 neurons \\
		\underline{Regularizer} : L1L2 (for layers 1 to 3) and none (for layer 4)\\
		\underline{Activation function} : ELU (for layers 1 to 3) and Sigmoid (layer 4) \\
		\underline{Initializer} : VarianceScaling, with uniform distribution \\
		(for layers 1 to 3) in fan\_in mode and \\ GlorotUniform distribution (layer 4) with null seed\\
		} \\
		\hline
	\end{tabular}}
	\caption{Architectures for the neural networks that the genetic algorithm finds for the scanned masses (600 to 1000 GeV) of the doublet VLL scenario.}
	\label{tab:table-VLBSM-EVO-2}
\end{table*}

\begin{table*}[ht!]    
	\centering
	\resizebox{0.91\textwidth}{!}{\begin{tabular}{|c|c|}
		\hline
    	Mass	& Singlet \\
		\hline
		\hline
		\midrule
		\makecell{100 GeV}  & \makecell{\underline{Layers} : 1 input + 1 hidden + 1 output. \\Hidden and input layers with 512 neurons each, \\ output layer with 3 neurons \\
		\underline{Regularizer} : L1L2 (for layers 1 to 4) and none (for layer 5)\\
		\underline{Activation function} : sigmoid (for all layers) \\
		\underline{Initializer} : RandomNormal, with 0 mean and 0.05 standard deviation \\
		(for layers 1 and 2) and VarianceScaling, with uniform \\ distribution in fan\_in mode (last layer) \\
		}  \\
		\hline
		\makecell{200 GeV} & \makecell{\underline{Layers} : 1 input + 1 hidden + 1 output. \\Hidden and input layers with 2048 neurons each, \\ output layer with 3 neurons \\
		\underline{Regularizer} : L1L2 (for layers 1 and 2) and none (for layer 3)\\
		\underline{Activation function} : sigmoid (for all layers)  \\
		\underline{Initializer} : VarianceScaling, with uniform distribution \\ (layer 1 and 2) in fan\_in mode and fan\_avg mode (last layer)\\
		} \\
		\hline
	\end{tabular}}
	\caption{Architectures for the neural networks that the genetic algorithm finds for the scanned masses (100 and 200 GeV) of the singlet VLL scenario.}
	\label{tab:table-VLBSM-EVO-3}
\end{table*}

\cleardoublepage

\section{ROC plots for Doublet VLLs}\label{sec:rocs}

\begin{figure}[htb!]
    \captionsetup{justification=raggedright,singlelinecheck=false}
	\subfloat{{\includegraphics[width=0.65\textwidth]{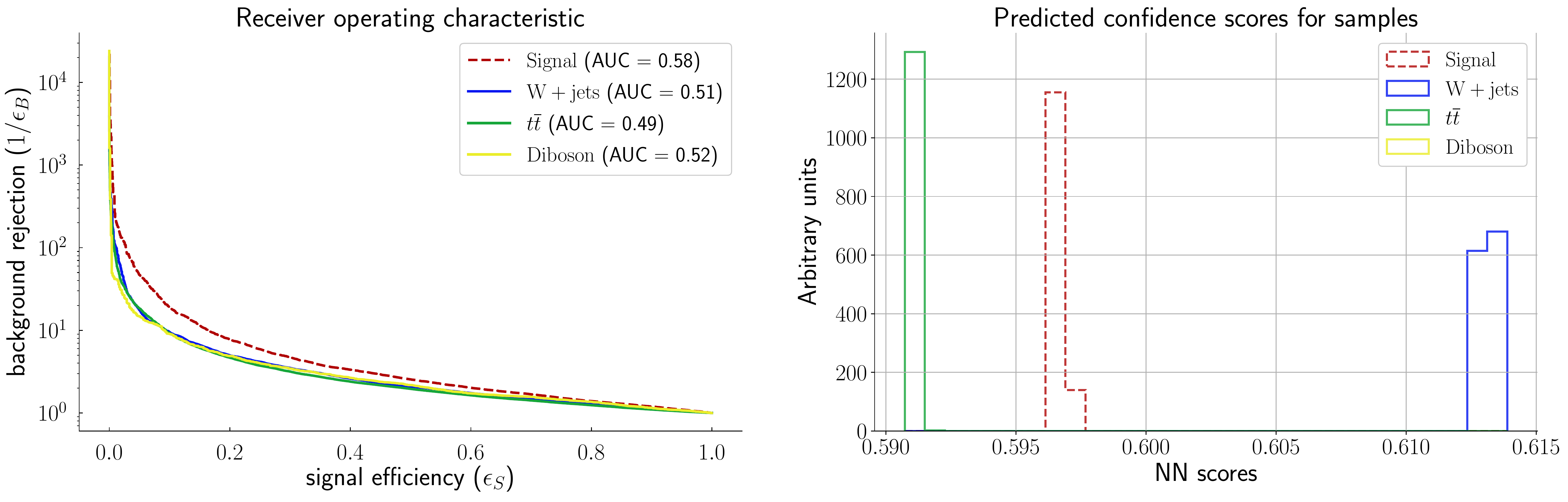} }} \\
	\subfloat{{\includegraphics[width=0.65\textwidth]{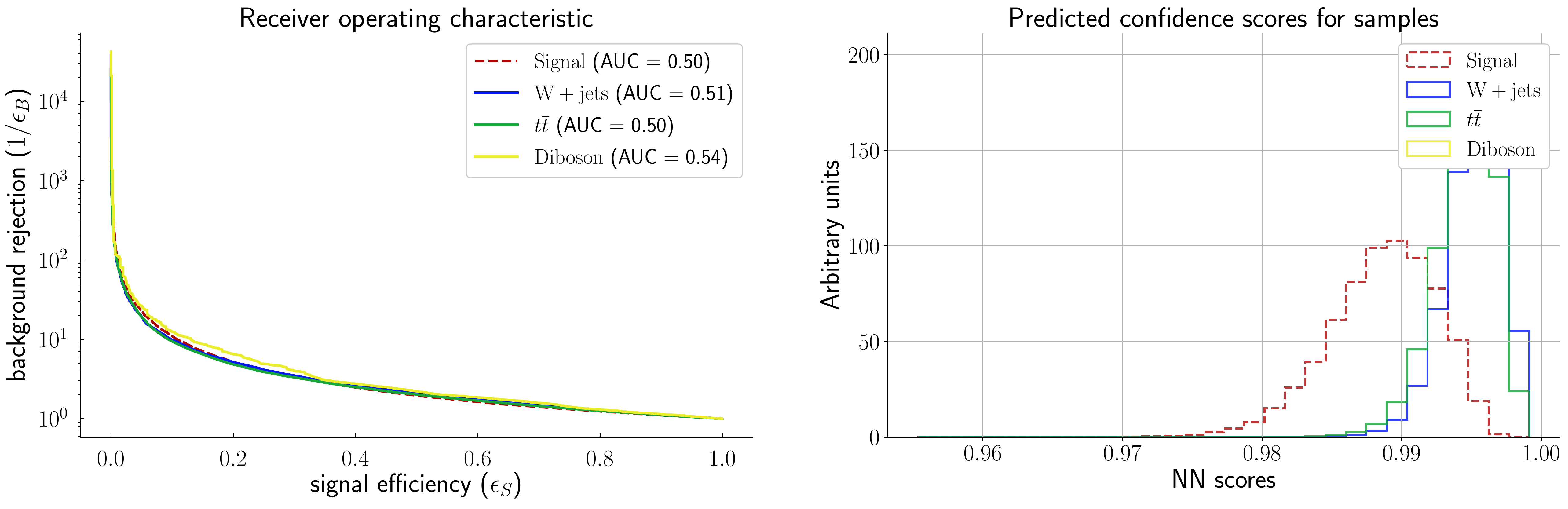} }} \\
	\subfloat{{\includegraphics[width=0.65\textwidth]{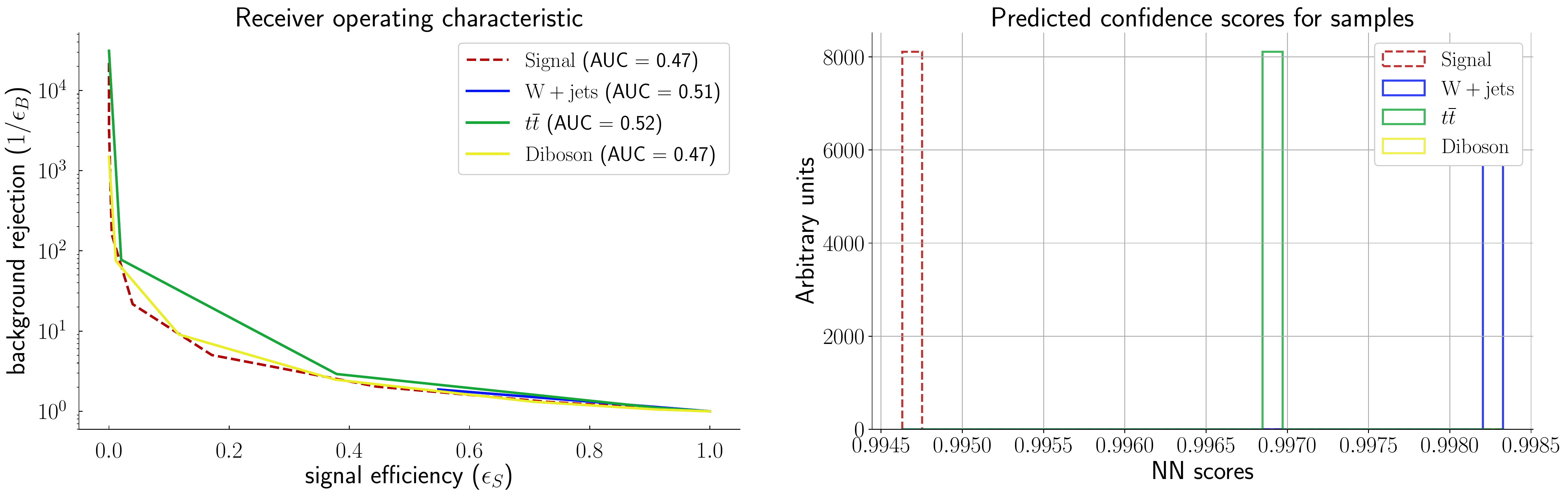} }} \\
	\subfloat{{\includegraphics[width=0.65\textwidth]{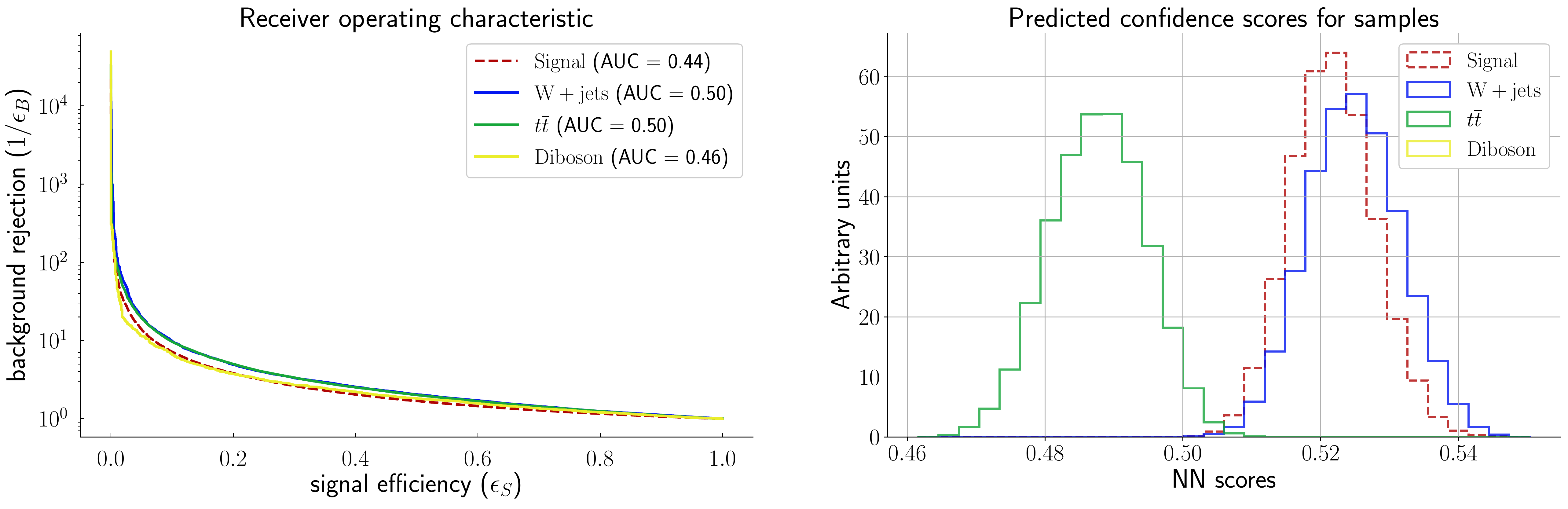} }} \\
	\subfloat{{\includegraphics[width=0.65\textwidth]{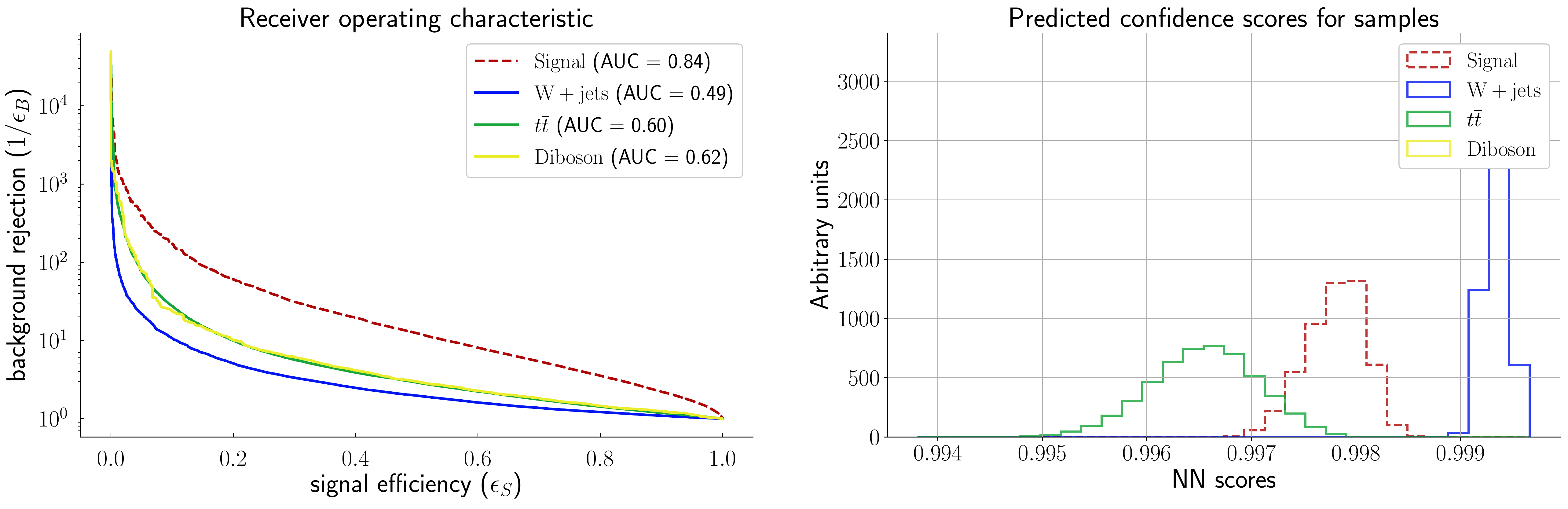} }} \\
	\caption{Receiver operator characteristic (ROC) plots for the neural networks of the doublet VLLs from the masses of 100 GeV (top) to 500 GeV (bottom).
		\label{fig:roc_plots}}
\end{figure}

\begin{figure}[htb!]
    \captionsetup{justification=raggedright,singlelinecheck=false}
	\subfloat{{\includegraphics[width=0.65\textwidth]{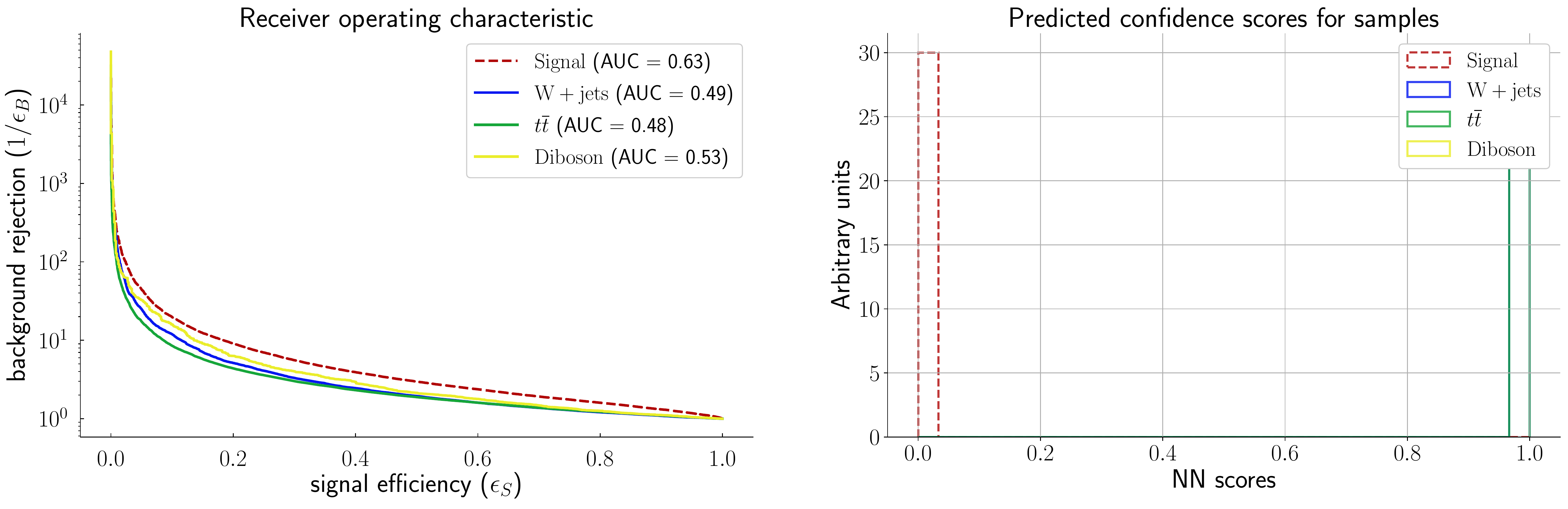} }} \\
	\subfloat{{\includegraphics[width=0.65\textwidth]{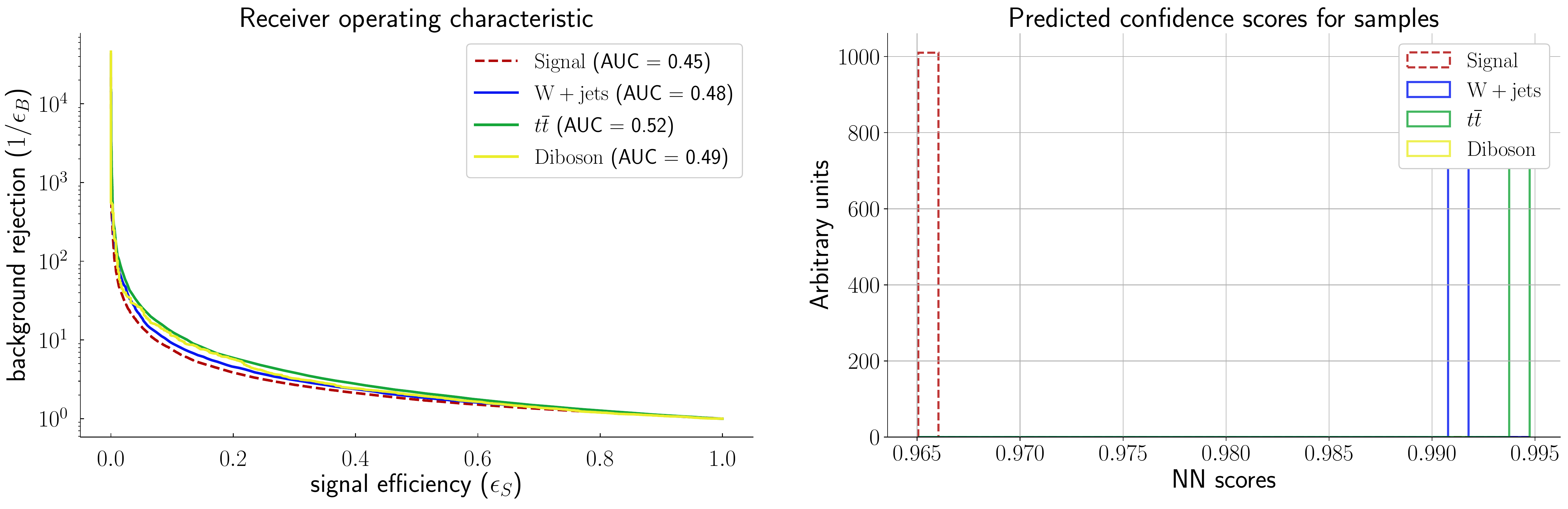} }} \\
	\subfloat{{\includegraphics[width=0.65\textwidth]{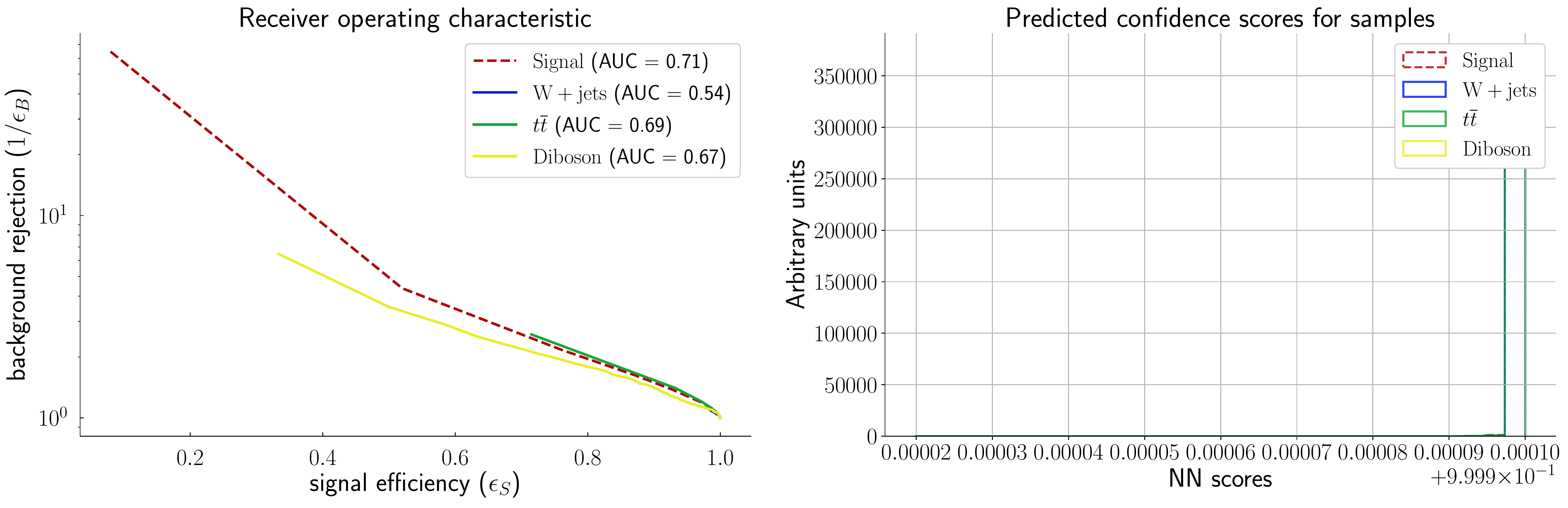} }} \\
	\subfloat{{\includegraphics[width=0.65\textwidth]{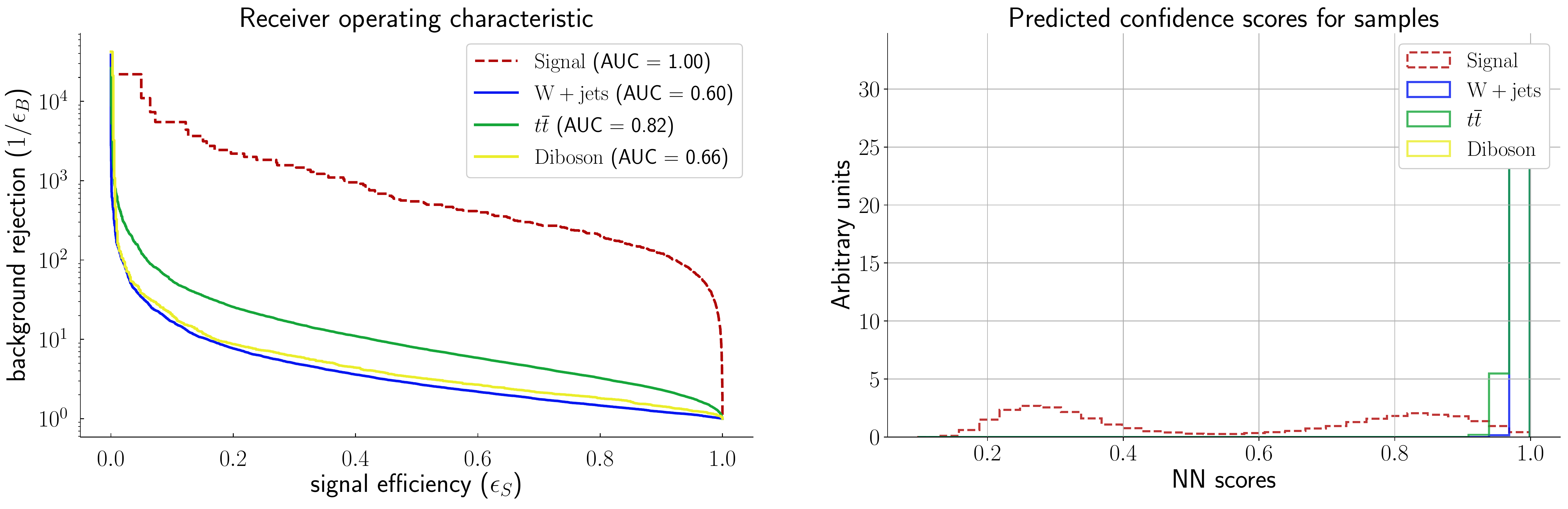} }} \\
	\subfloat{{\includegraphics[width=0.65\textwidth]{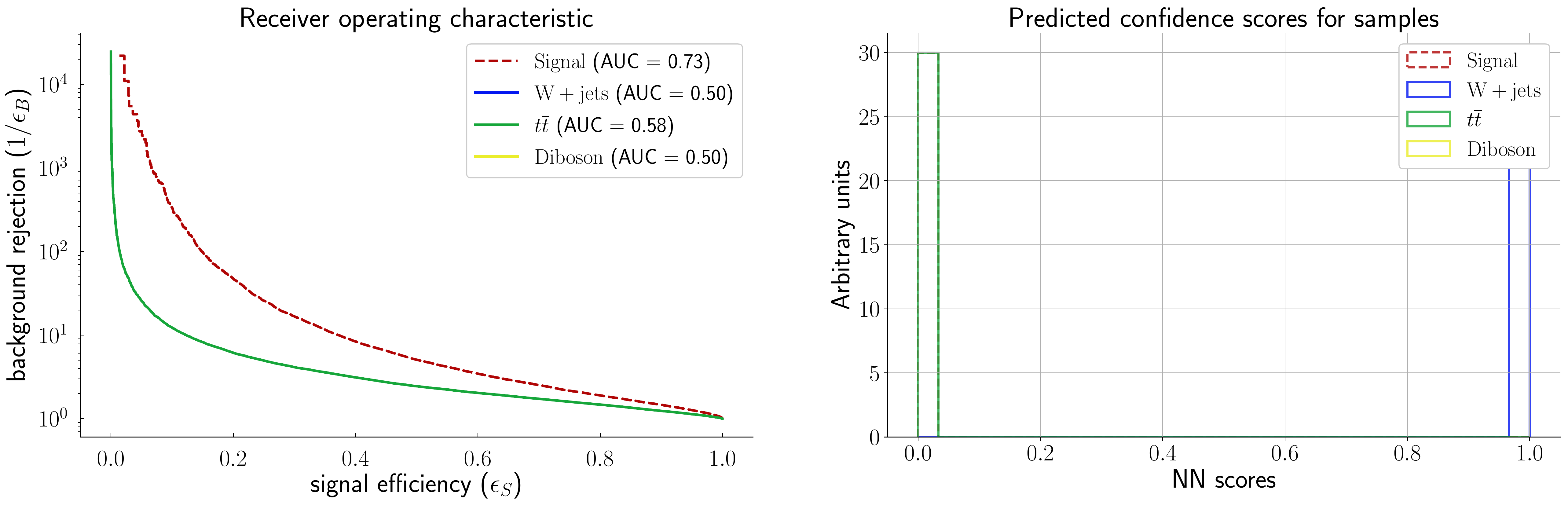} }} \\
	\caption{Receiver operator characteristic (ROC) plots for the neural networks of the doublet VLLs from the masses of 600 GeV (top) to 1000 GeV (bottom).
		\label{fig:roc_plots}}
\end{figure}

\bibliography{biblio}

\end{document}